\documentclass[journal]{IEEEtran}
\usepackage{cite,amssymb}
\usepackage[pdftex]{graphicx}
\usepackage{algorithm}
\usepackage[cmex10]{amsmath}
\usepackage{algorithm}
\usepackage{algorithm}
\usepackage{algpseudocode}
\usepackage{array,balance}

\ifCLASSOPTIONcompsoc
\usepackage[caption=false,font=normalsize,labelfont=sf,textfont=sf]{subfig}
\else
\usepackage[caption=false,font=footnotesize]{subfig}
\fi
\usepackage{url}
\usepackage{amsfonts}
\usepackage{exscale}
\usepackage{relsize}
\usepackage{multicol,color}
\newcommand{\mH}{\mathop{\rm H}}
\newcommand{\mT}{\mathop{\rm T}}
\begin{document}
\title{Pattern Synthesis via Complex-Coefficient Weight Vector Orthogonal Decomposition}
	\author{Xue Shi
	\thanks{The author is with the College of Computer Science and Cyber Security, Chengdu University of Technology, Chengdu 610059,
		China (e-mail: waterns@126.com).}}

\markboth{}
{Shell \MakeLowercase{\textit{et al.}}: Bare Demo of IEEEtran.cls for Journals}
\maketitle

\begin{abstract}
This paper presents a new array response control scheme named
complex-coefficient weight vector orthogonal decomposition
($ \textrm{C}^2\textrm{-WORD} $) and its application to
pattern synthesis.
The proposed $ \textrm{C}^2\textrm{-WORD} $ algorithm is a modified version of the existing WORD approach.
We extend WORD by allowing a complex-valued combining coefficient
in $ \textrm{C}^2\textrm{-WORD} $, and
find the optimal combining coefficient by maximizing white noise gain (WNG).
Our algorithm offers a closed-from expression to precisely control the array
response level of a given point starting from an arbitrarily-specified
weight vector.
In addition, it results less pattern variations on the uncontrolled angles.
Elaborate analysis shows that the proposed $ \textrm{C}^2\textrm{-WORD} $ scheme performs at least as good as the state-of-the-art
$\textrm{A}^\textrm{2}\textrm{RC} $ or WORD approach.
By applying $ \textrm{C}^2\textrm{-WORD} $ successively, we present a flexible and effective approach to pattern synthesis. 
Numerical examples are provided to demonstrate the flexibility and effectiveness of 
$ \textrm{C}^2\textrm{-WORD} $ in array response control as well as pattern synthesis.
\end{abstract}

\begin{IEEEkeywords}
Array pattern synthesis, array response control, white noise gain, array signal processing.
\end{IEEEkeywords}

\IEEEpeerreviewmaketitle

\section{Introduction}
\IEEEPARstart{A}{rray} antenna has found numerous applications to radar, navigation and wireless communication. 
Determining the complex weights for array elements to achieve a desired beampattern, i.e., array pattern synthesis, is a fundamental problem \cite{Er1992,newref3,2010-7,newref5,ref070}. 
For instance, in radar systems, it is desirable to mitigate returns from interfering signals by designing a pattern with several nulls at specified directions. In some communication systems, it is often required to shape multiple-beam patterns for multi-user reception. Additionally, synthesizing a pattern with broad mainlobe can extend monitoring areas in satellite remote sensing.

With regard to the problem of pattern synthesis, it is expected to control the sidelobe of array response to achieve a pencil beam pattern or to realize a shaped beam pattern complying to a given mask.
Quite a number of approaches to pattern synthesis have been reported
during the past several decades.
For example, given the mainlobe width,
Dolph provided an analytical solution to obtain a pattern with
minimum uniform sidelobe level \cite{ref01}. However, this method is only suitable
for arrays with particular geometries.
For arbitrary arrays, the global search approaches, such as
genetic algorithm (GA) \cite{ref002}, particle swarm optimization (PSO) \cite{ref05} and simulated annealing (SA) \cite{ref06}, are adopted to find the qualified weights
yielding satisfactory beampatterns. Nevertheless, the prohibitive amount of computation time 
limits their practical use.
Adaptive array theory \cite{refbookada1} has been utilized in \cite{ref07,nref07,ref10}
to synthesize desirable patterns. For this kind of method,
the array response levels are adjusted by assigning virtual
interferences. Note that the interference-to-noise ratios (INRs) of virtual interferences
are selected in an {\it{ad hoc}} manner, and how to determine
their parameters needs further investigation.

In the past few years, the convex optimization theory \cite{cvxbook} has been successfully exploited in pattern synthesis \cite{ref29,refaddcvx,newref6}. 
For example, authors of \cite{ref24} have shown how convex optimization can be utilized to design the optimal
pattern for arbitrary antenna arrays. Semidefinite programming (SDP) is employed in \cite{ref26} and \cite{ref299} to design nonuniform arrays with desired magnitude responses. 
Although most of the pattern synthesis problems have non-convex constraints,
the convex programming is also useful.
By iteratively linearizing the non-convex power pattern constraints, a series of convex subproblems are 
obtained in \cite{aps3} and further solved using second-order cone programming (SOCP). 
Authors of \cite{ref28} make the non-convex lower bound constraints on the beampattern
convexified, by exploiting the symmetric geometries of linear/planar arrays
and adopting a conjugate symmetric weight vector.
In \cite{ref27}, the semidefinite relaxation (SDR) technique \cite{snrf17} is employed to approximate the non-convex constraints in the pattern synthesis problem as convex.
Apart from the aforementioned methods, there also exist a few approaches attempting to synthesize patterns by utilizing the least-squares method \cite{ref30}, employing the Fast Fourier Transformation (FFT) \cite{ref33} or excitation matching approach \cite{newref7}.

More recently, we devised several flexible pattern synthesis approaches with array response control scheme. 
Starting from an arbitrarily-specified weight vector,
the accurate array response control (${\textrm{A}}^2{\textrm{RC}}$) algorithm \cite{refa2rc} synthesizes a desirable beampattern
by iteratively controlling the response of a single angle.
However, as reported in \cite{word}, the parameter determination
of ${\textrm{A}}^2{\textrm{RC}}$ is imperfect and the resulting
beampatterns may be distorted.
To overcome the drawback of ${\textrm{A}}^2{\textrm{RC}}$, 
the weight vector orthogonal decomposition (WORD) algorithm
is proposed in \cite{word}. The WORD based pattern synthesis
approach shares the same idea as that of ${\textrm{A}}^2{\textrm{RC}}$.
Nevertheless, different from ${\textrm{A}}^2{\textrm{RC}}$, 
the weight in WORD is orthogonally decomposed as 
two parts. By selecting the combining coefficient of the
resulting two orthogonal vectors in WORD, the response level at a 
single direction can be precisely adjusted, and a satisfactory
pattern can be synthesized.
In WORD algorithm, there are two real-valued candidates for the combining coefficient,
and the one that leads to a less pattern variation is selected.
In fact, as shall be presented later, one can realize the given response control task by adopting a complex-valued combining coefficient
in WORD scheme.
Since the candidate set of combining coefficients is
incomplete in the existing WORD algorithm, it may lead to performance loss
on the ultimate beampattern.

The drawbacks of WORD motivate us to consider a complex-coefficient weight vector orthogonal decomposition
($ \textrm{C}^2\textrm{-WORD} $) algorithm in this paper, where the
combining coefficient is allowed to be complex-valued but not limited to be real-valued.
Since the candidate set of weighting coefficients is extended, 
we can obtain a better performance than WORD in \cite{word}.
Moreover, we optimize the parameter of $ \textrm{C}^2\textrm{-WORD} $ by
maximizing the white noise gain (WNG) \cite{gain1,gain2,gain3} and
obtain a closed-form solution of the ultimate coefficient and weight vector.
In addition, we present a detailed analysis on the connection
between the proposed $ \textrm{C}^2\textrm{-WORD} $ and the existing
${\textrm{A}}^2{\textrm{RC}}$.
It is shown that $ \textrm{C}^2\textrm{-WORD} $ may degrade into
${\textrm{A}}^2{\textrm{RC}}$, but outperforms it under
general circumstances.
By applying $ \textrm{C}^2\textrm{-WORD} $ iteratively, a flexible and effective
array pattern synthesis approach is developed and its good performance
is validated under various situations.
Furthermore, by taking the steering vector uncertainty into consideration,
we will utilize $ \textrm{C}^2\textrm{-WORD} $ scheme
to realize robust sidelobe control and synthesis, which has
not been studied in neither ${\textrm{A}}^2{\textrm{RC}}$ nor WORD.

It should be mentioned that this paper focuses on the main concepts and fundamentals of 
$ \textrm{C}^2\textrm{-WORD} $ scheme, its application to robust sidelobe control and synthesis will be carried out in the
companion paper \cite{robust}.
This paper is organized as follows. In Section II, the problem formulation of array
response control,
${\textrm{A}}^2{\textrm{RC}}$ algorithm and WORD algorithm are briefly introduced. 
The $ \textrm{C}^2\textrm{-WORD} $ scheme is devised in Section III and its connection
with ${\textrm{A}}^2{\textrm{RC}}$ is discussed in Section IV.
The application of $ \textrm{C}^2\textrm{-WORD} $ to pattern synthesis is presented in Section V. In Section VI, we present numerical examples to demonstrate the performance of 
the proposed method. Conclusions are drawn in Section VII.

{\textit{Notations:}} We use bold upper-case and lower-case letters to represent
matrices and vectors, respectively.
In particular, we use $ {\bf I} $ to denote the identity matrix.
$ j\triangleq\sqrt{-1} $.
$ (\cdot)^{\mT} $, $ (\cdot)^* $
and $ (\cdot)^{\mH} $ stand for the transpose, complex conjugate and Hermitian
transpose, respectively.
$ |\cdot| $ denotes the absolute value and $ \|\cdot\|_2 $ denotes the $ l_2 $ norm.
$ {\bf v}(i) $ represents the $ i $th entry of vector $ {\bf v} $.
We use $ {\bf B}(i,l) $ to
stand for the element at the $ i $th row and $ l $th column of matrix $ {\bf B} $.
$ {\rm Diag}(\cdot) $ represents the diagonal matrix with the components of the input vector as the diagonal elements.
$ \Re(\cdot) $ and $ \Im(\cdot) $ denote the real
and imaginary parts, respectively.
$ {\rm det}(\cdot) $ is the determinant of a matrix.
$ \propto $ means direct proportion.
$ \mathbb{R} $ and $ \mathbb{C} $ denote the sets of all real and all complex numbers, respectively.
$ \mathcal{R}(\cdot) $ returns the column space
of the input matrix, and
$ \mathcal{R}^{\bot}(\cdot) $ is the orthogonal complementary space
of $ \mathcal{R}(\cdot) $.
$ {\bf P}_{\bf Z} $ and $ {\bf P}^{\bot}_{\bf Z} $ represent
the projection matrices onto $ \mathcal{R}({\bf Z}) $
and $ \mathcal{R}^{\bot}({\bf Z}) $, respectively.
$ \angle(\cdot) $ returns the argument of a complex number.
$ \oplus $ stands for the direct sum operator.

\section{Preliminaries}
\subsection{Response Control Formulation}
Without loss of generality and for the sake of clarity, we focus on herein the problem of one-dimensional 
response control. The extension to more complicated configurations is straight-forward.
First of all,  the array power response is expressed as
\begin{align}\label{modi0123}
L(\theta,\theta_0)={|{\bf w}^{\mH}{\bf a}(\theta)|^2}/{|{\bf w}^{\mH}{\bf a}(\theta_0)|^2}
\end{align}
where $ (\cdot)^{\mH} $ denotes the conjugate transpose, $ \bf{w} $ is the weight vector,
$ \theta_0 $ is the main beam axis,
$ {\bf a}(\theta) $ stands for the steering vector in direction $ \theta $.
More exactly, we have
\begin{align}
{\bf a}(\theta)=[ g_1(\theta)e^{-j\omega\tau_1(\theta)},\cdots,g_N(\theta)e^{-j\omega\tau_N(\theta)}]^{\mT}
\end{align}
where $ (\cdot)^{\mT} $ denotes the transpose operator,
$ {j}=\sqrt{-1} $ is the imaginary unit, $ \omega $ denotes the operating frequency,
$ g_n(\theta) $ denotes the pattern of the $n$th element, $ \tau_n(\theta) $ represents the time-delay between the $n$th element and the reference point, $ n=1,\cdots,N $.
Notice that the array response in \eqref{modi0123} is normalized by the power response at $ \theta_0 $ as commonly applied in practice.
The problem of array response control can thus be stated as: finding an appropriate weight vector which makes the normalized power response $ L(\theta,\theta_0) $ meet  specific requirements.

\subsection{$ \textrm{A}^2\textrm{RC} $ Algorithm}
Recently, an accurate array response control ($ {\textrm {A}}^2\textrm{RC} $) algorithm
has been developed in \cite{refa2rc} to precisely control the response level at one given point.
For a given previous weight vector $ {\bf w}_{k-1} $ and an angle $ \theta_{k} $
to be controlled,
the weight vector of ${\textrm A}^2{\textrm{RC}}$ is updated as 
\begin{align}\label{eqna2rc}
{\bf w}_{k}={\bf w}_{k-1}+\mu_{k}{\bf a}(\theta_{k})
\end{align}
where $ k $ represents the index of step, $ \mu_{k} $ is the only parameter that can be determined
by the desired level $ \rho_k $ at $ \theta_k $.
More specifically, 
to realize the array response control
task at $ \theta_k $, i.e., 
$ L_{k}(\theta_k,\theta_0)=\rho_k $,
where
\begin{align}
L_{k}(\theta,\theta_0)=
{| {\bf{w}}^{\mH}_{k}{\bf{a}}(\theta) |^2}\big/
{| {\bf{w}}^{\mH}_{k}{\bf{a}}(\theta_0) |^2}
\end{align}
it has been shown in \cite{refa2rc} that $ \mu_{k} $ locates in a 
circle:
\begin{align}
\mathbb{C}_{{\mu}_k}=\left\{
{\mu}_k\Big|
\big\|[	{\Re}({\mu}_k),{\Im}({\mu}_k)]^{\mT}-
{\bf c}_{\mu_k}\big\|_2=R_{\mu_k}
\right\}
\end{align}	
with center $ {\bf c}_{\mu} = 
[-{\Re}\left({\bf Q}_k(1,2)\right), {\Im}\left({\bf Q}_k(1,2)\right)]^{\mT}/{\bf Q}_k(2,2) $
and radius $ R_{\mu}={\sqrt{-{\rm det}({\bf Q}_k)}}/{|{\bf Q}_k(2,2)|} $,
where $ {\bf Q}_k=
[{\bf w}_{k-1},
{\bf a}(\theta_k)]^{\mH}
({\bf a}(\theta_k) {\bf a}^{\mH}(\theta_k)-\rho_k{\bf a}(\theta_0){\bf a}^{\mH}(\theta_0))
[{\bf w}_{k-1},
{\bf a}(\theta_k)] $.

In addition, to generate a less pattern distortion, the optimal $ {\mu}_{k,\star} $ is experimentally selected in \cite{refa2rc} as
\begin{align}\label{e2156}
{\mu}_{k,\star}={\mu_{k,s}}\triangleq{\rm arg}~\min_{{\mu_k}\in{\mathbb{C}_{{\mu}_k}}}~|\mu_k|.
\end{align}
However, a satisfactory performance may not be always
guaranteed for ${\textrm A}^2{\textrm{RC}}$, mainly due to its
imperfect parameter determination scheme.

\subsection{$ \textrm{WORD} $ Algorithm}
To alleviate the drawback of ${\textrm A}^2{\textrm{RC}}$,
a novel weight vector orthogonal decomposition (WORD) algorithm 
was presented in \cite{word} on the foundation of adaptive array theory.
More specifically, 
for a given weight vector $ {\bf w}_{k-1} $,
an angle $ \theta_k $ to be controlled and its desired level $ \rho_k $,
WORD algorithm updates its weight vector as
\begin{align}\label{tap0004}
{\bf{w}}_{k}=\begin{bmatrix}
{\bf{w}}_{k-1,\bot}&
{\bf{w}}_{k-1,\Arrowvert}
\end{bmatrix}
\begin{bmatrix}
1& {\beta}_{k}
\end{bmatrix}^{\mT},~\beta_k\in\mathbb{R}
\end{align}
where $ {\bf{w}}_{k-1,\bot}$ and $ {\bf{w}}_{k-1,\Arrowvert}$ are defined 
as
\begin{align}\label{word01}
	{\bf{w}}_{k-1,\bot}\triangleq{\bf{P}}^{\bot}_{[{\bf{a}}(\theta_{k})]}{\bf{w}}_{k-1},~~
	{\bf{w}}_{k-1,\Arrowvert}\triangleq{\bf{P}}_{[{\bf{a}}(\theta_{k})]}{\bf{w}}_{k-1}
\end{align}
with $ k $ denoting the step index.
In \eqref{tap0004}, the real-valued $ {\beta}_k $ can be selected to be either
$ {\beta}_a $ or $ {\beta}_b $, both of which can be determined by the
desired level $ \rho_k $ at $ \theta_k $. In \cite{word}, it has been derived that
\begin{align}\label{close003}
{\beta}_a=
\dfrac{-{\Re}({\bf B}_k(1,2))+
	d}{{\bf B}_k(2,2)},~~~
{\beta}_b=\dfrac{-{\Re}({\bf B}_k(1,2))-
	d}{{\bf B}_k(2,2)}
\end{align}
where $ {\bf B}_k $ and $ d $ satisfy
\begin{align}
\label{close004}\!\!\!{\bf B}_k\!&=\!\!\begin{bmatrix}
{\bf{w}}^{\mH}_{\bot}{\bf{a}}(\theta_k)\\ {\bf{w}}^{\mH}_{\Arrowvert}{\bf{a}}(\theta_k)
\end{bmatrix}\!\!
\begin{bmatrix}
{\bf{w}}^{\mH}_{\bot}{\bf{a}}(\theta_k)\\ {\bf{w}}^{\mH}_{\Arrowvert}{\bf{a}}(\theta_k)
\end{bmatrix}^{\mH}\!\!\!\!\!-\!
\rho_k\!\!
\begin{bmatrix}
{\bf{w}}^{\mH}_{\bot}{\bf{a}}(\theta_0)\\ {\bf{w}}^{\mH}_{\Arrowvert}{\bf{a}}(\theta_0)
\end{bmatrix}\!\!
\begin{bmatrix}
{\bf{w}}^{\mH}_{\bot}{\bf{a}}(\theta_0)\\ {\bf{w}}^{\mH}_{\Arrowvert}{\bf{a}}(\theta_0)
\end{bmatrix}^{\mH}\!\!\!\\
\label{close005}&~~~~~~d=\sqrt{{\Re}^2({\bf B}_k(1,2))-{\bf B}_k(1,1){\bf B}_k(2,2)}.
\end{align}
In \eqref{close004}, $ {\bf{w}}_{\bot}$ and $ {\bf{w}}_{\Arrowvert}$
are the short notations of $ {\bf{w}}_{k-1,\bot}$ and $ {\bf{w}}_{k-1,\Arrowvert}$,
respectively, and will be used in our later discussions.
To obtain the ultimate expression of $ {\bf w}_{k} $ that adjusts
the response level of $ \theta_k $ to $ \rho_k $,
the one (either ${\beta}_a$ or ${\beta}_b$) that
minimizes $ F({\beta})=\| {\bf P}^{\bot}_{{{\bf{w}}_{k-1}}}{{\bf{w}}_{k}}/
{\|{\bf{w}}_{k} \|_2} \|^2_2 $
is selected.

\section{Array Response Control via $ \textrm{C}^2\textrm{-WORD} $}
In this section, we present a new complex-coefficient weight vector orthogonal decomposition ($ \textrm{C}^2\textrm{-WORD} $) algorithm
by modifying the WORD algorithm in \cite{word}.

\subsection{$ \textrm{C}^2\textrm{-WORD} $ Algorithm}
Before presenting the proposed $ \textrm{C}^2\textrm{-WORD} $ algorithm, we
first provide the following
Lemma.
\newtheorem{theoreml}{Lemma}
\begin{theoreml}
	For $ \forall\beta_k\in\mathbb{R}$ and the corresponding $ {\bf{w}}_{k} $ in \eqref{tap0004}, 
	there exists $ \widetilde{\beta}_k\in\mathbb{C} $ and 
	a corresponding
	$ \widetilde{{\bf w}}_{k} $ satisfying
	\begin{align}\label{eqn1023}
	\widetilde{{\bf w}}_{k}=
	\begin{bmatrix}{\bf{w}}_{\bot}&{\bf{w}}_{\Arrowvert}\end{bmatrix}
	\begin{bmatrix}	1& \widetilde{\beta}_k \end{bmatrix}^{\mT}
	\end{align}
	such that
\begin{align}\label{eq12}
L_{k}(\theta_k,\theta_0)\big|_{{\bf w}=\widetilde{{\bf w}}_{k}}=L_{k}(\theta_k,\theta_0)\big|_{{\bf w}={{\bf w}}_{k}}.
\end{align}
Moreover, $ \widetilde{\beta}_k $
is non-trivially complex-valued (i.e.,
$ \widetilde{\beta}_k\notin\mathbb{R} $)
in most cases.
\end{theoreml}
\begin{IEEEproof}
	See Appendix A.
\end{IEEEproof}

Lemma 1 implies that there exists complex-valued $ \widetilde{\beta}_k $ leading to the same response level at $ \theta_k $ as that of the prescribed real-valued $ {\beta}_k $. 
Therefore, it is more reasonable to assign a complex-valued $ \beta_k $
in the WORD scheme. This leads to
the complex-coefficient weight vector orthogonal decomposition ($ \textrm{C}^2\textrm{-WORD} $) algorithm as presented next.

More specifically, given the previous weight vector $ {\bf{w}}_{k-1} $,
in order to adjust the array response level of $ \theta_k $
to its desired level $ \rho_k $, we propose to update the
weight vector as
\begin{align}\label{tap1005}
{\bf{w}}_{k}=\begin{bmatrix}
{\bf{w}}_{\bot}&
{\bf{w}}_{\Arrowvert}
\end{bmatrix}
\begin{bmatrix}
1& {\beta}_{k}
\end{bmatrix}^{\mT},~\beta_k\in\mathbb{C}.
\end{align}
Different from the weight vector update of $ \textrm{WORD} $ in Eqn. \eqref{tap0004}, 
the parameter $ \beta_k $ in \eqref{tap1005} is complex-valued but not limited
to be real-valued, although we have designated an identical notation (i.e., $ \beta_k $).
We will present the benefits later.

To realize the array response control task at $ \theta_k $, i.e.,
\begin{align}\label{word005}
L_{k}(\theta_k,\theta_0)=\rho_k
\end{align}
we first find the trajectory of all possible $ \beta_{k} $'s in \eqref{tap1005}.
To do so, we substitute \eqref{tap1005} into \eqref{word005}
and obtain 
$ {\bf{z}}^{\mH}_{k}{\bf{B}}_{k}{\bf{z}}_{k}=0 $,
where $ {\bf z}_{k} \triangleq [1~~\beta_{k}]^{\mT} $, $ {\bf{B}}_{k} $ is a $ 2\times2 $ Hermitian matrix given in \eqref{close004}.
After some calculation, we can obtain the following proposition
that describes the geometrical distribution of $ \beta_k $.

\newtheorem{theorem}{Proposition}
\begin{theorem}
	Suppose that $ {\beta}_k $ is a solution of \eqref{tap1005} and \eqref{word005}, it can be
	derived that the trajectory set of $ {\beta}_k $ is a 
	circle $ \mathbb{C}_{{\beta}_k} $:
	\begin{align}\label{setCbeta}
	\mathbb{C}_{{\beta}_k}=\left\{
	{\beta}_k\Big|
	\big\|[	{\Re}({\beta}_k),{\Im}({\beta}_k)]^{\mT}-
	{\bf c}_{\beta_k}\big\|_2=R_{\beta_k}
	\right\}
	\end{align}	
	with center
	\begin{align}\label{eqn0030}
	{\bf c}_{\beta_k}=\dfrac{1}{{\bf B}_k(2,2)} 
	\begin{bmatrix}-{\Re}\left({\bf B}_k(1,2)\right)\\ {\Im}\left({\bf B}_k(1,2)\right)\end{bmatrix}
	\end{align}
	and radius
	\begin{align}\label{eqn0031}
	R_{\beta_k}={\sqrt{-{\rm det}({\bf B}_k)}}\big/{|{\bf B}_k(2,2)|}.
	\end{align}
\end{theorem}
\begin{IEEEproof}
	See Appendix B.
\end{IEEEproof}

\begin{figure}[!t]
	\centering
	\includegraphics[width=3.1in]{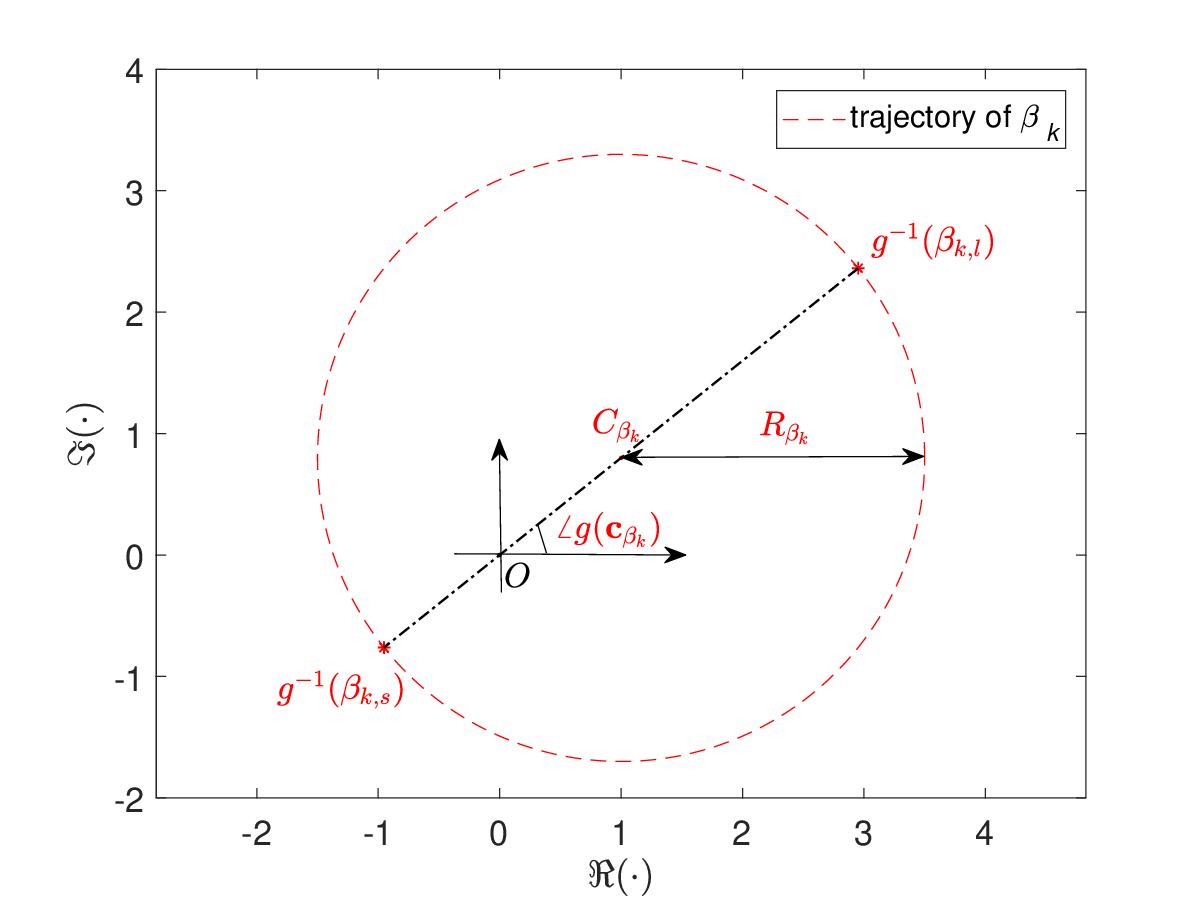}
	\caption{Geometrical distribution of $ \beta_k $.}
	\label{tspcirclebetabasic}
\end{figure}

Fig. \ref{tspcirclebetabasic} presents a geometric interpretation of Proposition 1.
It can be readily known that the existing WORD algorithm selects
the parameter $ \beta_k $ from the real-valued elements of $ {\mathbb{C}_{{\beta}_k}} $.
For our $ \textrm{C}^2\textrm{-WORD} $ algorithm, we have extended
the feasible set to complex domain.
By doing so, the resulting performance may be improved,
since it's possible to select a more appropriate $ \beta_k $ that
may not be real-valued.

\textit{Remark 1}: Proposition 1 is valid only if $ {\bf B}_k(2,2)\neq0 $, which is 
equivalent to
$ \rho_k| {\bf{w}}^{\mH}_{\Arrowvert}{\bf{a}}(\theta_0) |^2\neq| {\bf{w}}^{\mH}_{\Arrowvert}{\bf{a}}(\theta_k) |^2 $
or simply $ {\rho_k}|{\bf a}^{\mH}(\theta_k){\bf a}(\theta_0)|^2\neq
|{\bf a}^{\mH}(\theta_k){\bf a}(\theta_k)|^2 $.
As a matter of fact, if $ {\rho_k}|{\bf a}^{\mH}(\theta_k){\bf a}(\theta_0)|^2=
|{\bf a}^{\mH}(\theta_k){\bf a}(\theta_k)|^2 $, one obtains
\begin{align}\label{rho1}
\rho_k=|{\bf a}^{\mH}(\theta_k){\bf a}(\theta_k)|^2/|{\bf a}^{\mH}(\theta_k){\bf a}(\theta_0)|^2
\triangleq{\widetilde{\rho}_k}.
\end{align} 
Clearly, $ {\widetilde{\rho}_k} $ denotes the array response level at $ \theta_k $ when taking the weight vector as 
$ {\bf a}(\theta_k) $. Note that the response is normalized by the output at $ \theta_0 $, but its beam axis steers to $ \theta_k\neq\theta_0 $. Therefore, we have
$ {\widetilde{\rho}_k}>1 $ under normal circumstances. 
For this reason, in the following discussions we reasonably assume that $ {\rho_k}\in[0,1] $, which makes $ {\rho_k}\neq{\widetilde{\rho}_k} $ or $ {\bf B}_k(2,2)\neq0 $
easily satisfied.


\subsection{Selection of $ \beta_k $}
Proposition 1 indicates that
there exist infinitely many solutions of (complex-valued) $ \beta_k $ adjusting
the array response level of $ \theta_k $ to its desired value $ \rho_k $.
Then, an optimal one, denoted as $ \beta_{k,\star} $, should be selected
to further finish the array response control task \eqref{word005}.

In this paper,
we optimize the parameter $ \beta_k $ by maximizing 
the white noise gain (WNG)
\cite{gain1,gain2,gain3},
which is denoted as $ G $ and satisfies
\begin{align}\label{qua501}
G=\dfrac{|{\bf w}^{\mH}{\bf a}(\theta_0)|^2}
{\|{\bf w}\|^2_2}.
\end{align}
The WNG measures the signal-to-noise ratio (SNR) improvement,
i.e., the ratio of the SNR at the output to the SNR at 
the input, in the white noise scenario.
Combining \eqref{tap1005} and Proposition 1,
the following constrained problem can be formulated
to determine the ultimate $ \beta_{k,\star} $:
\begin{subequations}\label{criterion}
	\begin{align}
	\label{mini_obj}\max_{\beta_k}&~~~G_k=\dfrac{|{\bf w}^{\mH}_{k}{\bf a}(\theta_0)|^2}
	{\|{\bf w}_{k}\|^2_2}\\
	\label{criterion4}{\rm subject~to}&~~~{\bf{w}}_{k}=\begin{bmatrix}
	{\bf{w}}_{\bot}&{\bf{w}}_{\Arrowvert}\end{bmatrix}
	\begin{bmatrix}1& {\beta}_k\end{bmatrix}^{\mT}\\
	\label{criterion3}&~~~{\beta}_k\in{\mathbb{C}}_{{\beta}_k}
	\end{align}
\end{subequations}
where $ \mathbb{C}_{{\beta}_k} $ is given in \eqref{setCbeta}.
Although problem \eqref{criterion} is nonconvex,
its optimal solution can be analytically expressed
as the following proposition described.
\begin{theorem}
	The optimal solution $ \beta_{k,\star} $ of problem \eqref{criterion} is given by
\begin{align}\label{e2126}
\!\!\!\!\!\!\beta_{k,\star}={\beta_{k,l}}\triangleq{\rm arg}~\max_{{\beta_k}\in{\mathbb{C}_{{\beta}_k}}}\!|\beta_k|
=(|{\bf c}_{\beta_k}|+{R_{\beta_k}})e^{j\angle g({\bf c}_{\beta_k})}
\end{align}		
where $ g(\cdot) $ is a function satisfying $ g({\bf c})={\bf c}(1)+j{\bf c}(2) $
for a two-dimensional input vector.
\end{theorem}
\begin{IEEEproof}
	See Appendix C.	
\end{IEEEproof}

From Proposition 2, we learn that the optimal $ \beta_{k,\star} $ has
maximum modulo among all $ {\beta}_k $'s in $ {\mathbb{C}_{{\beta}_k}} $, refer to
$ g^{-1}({\beta_{k,l}}) $ in
Fig. \ref{tspcirclebetabasic} for its location.
Once the optimal $ \beta_{k,\star} $ is determined,
we can express the ultimate  weight vector of $ \textrm{C}^2\textrm{-WORD} $ as
\begin{align}\label{e06}
{\bf{w}}_{k}=\begin{bmatrix}
{\bf{w}}_{\bot}&{\bf{w}}_{\Arrowvert}\end{bmatrix}
\begin{bmatrix}1& {\beta}_{k,\star}\end{bmatrix}^{\mT}.
\end{align}

In addition, we can derive the following interesting result about
the obtained $ {\bf{w}}_{k} $ in \eqref{e06}.

\begin{theorem}
Taking $ {\bf w}_{k-1}={\bf a}(\theta_0) $ and assuming that
$ {{\bf a}^{\mH}(\theta_k){\bf a}(\theta_0)}\neq 0 $,
	then the resulting $ {\bf{w}}_{k} $ in \eqref{e06}
	is the global optimal solution of the following problem, i.e.,
	\begin{subequations}\label{eqn011}
		\begin{align}
		\label{eqn0102}
		\max_{{\bf w}\in{\mathbb{C}}^N}&~~~\dfrac{|{\bf w}^{\mH}{\bf a}(\theta_0)|^2}
		{\|{\bf w}\|^2_2}\\
		\label{eqn012}
		{\rm subject~to}&~~~{|{\bf w}^{\mH}{\bf a}(\theta_k)|^2}/
		{|{\bf w}^{\mH}{\bf a}(\theta_0)|^2}={\rho_k}.
		\end{align}
	\end{subequations}
	\begin{IEEEproof}
		See Appendix D.
	\end{IEEEproof}
\end{theorem}

According to Proposition 3,
the resulting weight vector 
of $ \textrm{C}^2\textrm{-WORD} $ is optimal in maximizing WNG with the
response level constraint \eqref{eqn012}, provided
that $ {\bf a}(\theta_0) $ is taken as the previous weight and 
$ {\bf a}(\theta_k) $ is non-orthogonal to $ {\bf a}(\theta_0) $.
It should be noted that the constraint of orthogonal decomposition (i.e., \eqref{criterion4}) is not assigned in problem \eqref{eqn011}.
Since $ \beta_{k,\star} $ may be complex-valued,
the above result may not be true for the traditional WORD algorithm.
Finally, we summarize the proposed $ \textrm{C}^2\textrm{-WORD} $ algorithm
in Algorithm \ref{fadsct33orps}.

\begin{algorithm}[!t]
	\caption{$ \textrm{C}^2\textrm{-WORD} $ Algorithm}\label{fadsct33orps}
	\begin{algorithmic}[1]
		\State prescribe beam axis $ \theta_0 $ and index $ k $, give the previous weight vector $ {\bf w}_{k-1} $, 
		direction $ {\theta_{k}} $ and the corresponding desired level
		$ {\rho_{k}} $
		\State determine the optimal $ \beta_{k,\star} $ in \eqref{e2126}
		\State output the new weight vector $ {\bf w}_k $ in \eqref{e06}
	\end{algorithmic}
\end{algorithm}

\section{Connection Between $ \textrm{C}^2\textrm{-WORD} $ and 
	$ \textrm{A}^2\textrm{RC} $}
In the preceding section, we extend WORD algorithm \cite{word} by allowing
the complex-valued coefficient in $ \textrm{C}^2\textrm{-WORD} $.	
Since $ \textrm{C}^2\textrm{-WORD} $ optimizes its parameter in a 
larger range, a more satisfactory performance can be obtained comparing
to WORD.
To have a better understanding,
we next present the connection between $ \textrm{C}^2\textrm{-WORD} $ and 
the existing ${\textrm A}^2{\textrm{RC}}$
in \cite{refa2rc}.
It is shown that ${\textrm A}^2{\textrm{RC}}$, WORD and
$ \textrm{C}^2\textrm{-WORD} $ may obtain identical results under
specific circumstances.

To begin with, we re-express the weight vector update of $ \textrm{C}^2\textrm{-WORD} $,
i.e., Eqn. \eqref{tap1005}, as
\begin{align}\label{e2127}
{\bf{w}}_{k}&=\begin{bmatrix}{\bf{w}}_{\bot}&{\bf{w}}_{\Arrowvert}\end{bmatrix}
\begin{bmatrix}	1& {\beta}_k\end{bmatrix}^{\mT}\nonumber\\
&={\bf{P}}^{\bot}_{[{\bf{a}}(\theta_{k})]}{\bf{w}}_{k-1}+
{\beta}_k{\bf{P}}_{[{\bf{a}}(\theta_{k})]}{\bf{w}}_{k-1}
\nonumber\\
&=\left(
{\bf I}-\frac{{\bf a}(\theta_k){\bf a}^{\mH}(\theta_k)}{{\bf a}^{\mH}(\theta_k){\bf a}(\theta_k)}
\right){\bf{w}}_{k-1}+
\frac{{\beta}_k{\bf a}(\theta_k){\bf a}^{\mH}(\theta_k)}{{\bf a}^{\mH}(\theta_k){\bf a}(\theta_k)}{\bf{w}}_{k-1}
\nonumber\\
&={\bf{w}}_{k-1}+T(\beta_k)\cdot{\bf a}(\theta_k)
\end{align}
where the transformation $ T(\beta_k) $ is defined as
\begin{align}\label{e2128}
T(\beta_k)\triangleq\dfrac{(\beta_k-1){\bf a}^{\mH}(\theta_k){\bf{w}}_{k-1}}{{\bf a}^{\mH}(\theta_k){\bf a}(\theta_k)}.
\end{align}
Comparing \eqref{e2127} with \eqref{eqna2rc}, one can see that
the weight vectors
of $ \textrm{C}^2\textrm{-WORD} $ and ${\textrm A}^2{\textrm{RC}}$ are
updated in the same forms, while the difference relies on the parameter selections,
see \eqref{e2126} and \eqref{e2156}, respectively.
In addition,
we notice that
the component in $ \mathcal{R}^{\bot}({\bf a}(\theta_k)) $
has no contribution to the array response level at $ \theta_k $,
but may affect the responses of the uncontrolled points.
From \eqref{e2127},
we know that no redundant component has been added into the previous
weight $ {\bf w}_{k-1} $ for $ \textrm{C}^2\textrm{-WORD} $ algorithm.
The resulting benefit may be the less pattern variations at the
uncontrolled region.

In fact, the one-one mapping $ T(\cdot) $ can transform $ \beta_k $ of
$ \textrm{C}^2\textrm{-WORD} $ into $ \mu_k $ of ${\textrm A}^2{\textrm{RC}}$, and vice versa. 
Moreover, it's not hard to find that 
\begin{align}
\left\{
T(\beta_k)\big|L(\theta_k,\theta_0)=\rho_k
\right\}=
\left\{
T(\beta_k)\big|\beta_k\in{\mathbb{C}_{{\beta}_k}}
\right\}={\mathbb{C}}_{\mu_k}.\nonumber
\end{align}
In addition, $ \textrm{C}^2\textrm{-WORD} $ obtains
the same result as ${\textrm A}^2{\textrm{RC}}$
under certain circumstances. 
To see this, we first define
\begin{align}\label{e2137}
{\mu_{k,l}}\triangleq{\rm arg}~\max_{{\mu_k}\in{\mathbb{C}_{\mu_k}}}~|\mu_k|.
\end{align}
Then, the following proposition can be established.
\begin{theorem}
	For the given $ {\bf{w}}_{k-1} $, $ \theta_k $ and $ {\rho}_k $, 
	we have
	$ T({\beta}_{k,\star})={\mu}_{k,s}={\mu}_{k,\star} $ if and only if
	\begin{align}\label{e2145}
	{\bf c}_{\beta_k}(2)=0~~{\rm and}~~
	{\bf c}_{\beta_k}(1)\in[0,1]
	\end{align}
	where $ {\bf c}_{\beta_k} $ is given in \eqref{eqn0030}.
	Meanwhile, $ T({\beta}_{k,\star})={\mu}_{k,l} $ if and only if
	\begin{align}\label{e2146}
	{\bf c}_{\beta_k}(2)=0~~{\rm and}~~
	{\bf c}_{\beta_k}(1)\in(-\infty,0)\cup(1,+\infty).
	\end{align}
	Additionally, if $ T({\beta}_{k,\star})\in\{{\mu}_{k,s},{\mu}_{k,l}\} $, we have 
	$ {\beta}_{k,\star}\in\mathbb{R} $.
\end{theorem}
\begin{IEEEproof}
	See Appendix E.
\end{IEEEproof}

According to Proposition 4, we know that
$ \textrm{C}^2\textrm{-WORD} $ results an identical result
as ${\textrm A}^2{\textrm{RC}}$, provided that \eqref{e2145}
is satisfied.
Otherwise, it can be predicted that 
$ \textrm{C}^2\textrm{-WORD} $ outperforms
${\textrm A}^2{\textrm{RC}}$ (in the sense of WNG).
Moreover, note that the parameter $ {\mu}_{k,l} $, which is defined
in \eqref{e2137} and
is never selected in ${\textrm A}^2{\textrm{RC}}$,
may correspond to the optimal $ {\beta_{k,\star}} $ of 
$ \textrm{C}^2\textrm{-WORD} $.
In addition, it can be readily find from
Proposition 4 that $ \textrm{C}^2\textrm{-WORD} $ may
obtain the same result as that of WORD, provided that $ {\bf c}_{\beta_k}(2)=0 $.

Proposition 4 summarizes the condition 
when $ \textrm{C}^2\textrm{-WORD} $ and
${\textrm A}^2{\textrm{RC}}$ becomes equivalent. 
However, it is still unclear in which concrete cases
these two algorithms result an identical weight vector.
Before answering this question, we define
\begin{subequations}
	\begin{align}
	\label{rho3}\overline{{\rho}}_k&\triangleq\dfrac{| {\bf{a}}^{\mH}(\theta_k){\bf{a}}(\theta_k) |^2}
	{| {\bf{a}}^{\mH}(\theta_0){\bf{a}}(\theta_0) |^2}=
	\dfrac{\|{\bf a}(\theta_k)\|^4_2}{\|{\bf a}(\theta_0)\|^4_2}\\
	\label{rho2}{\breve{\rho}_k}&\triangleq
	\dfrac{|{\bf{w}}^{\mH}_{k-1}{\bf a}(\theta_k){\bf a}^{\mH}(\theta_k){\bf a}(\theta_k)|}
	{|{\bf{w}}^{\mH}_{k-1}{\bf a}(\theta_0){\bf a}^{\mH}(\theta_0){\bf a}(\theta_k)|}.
	\end{align}
\end{subequations}
Then, the following two corollaries of Proposition 4 can be obtained.
\newtheorem{theorem2}{Corollary}
\begin{theorem2}
	If the previous weight vector is set as $ {\bf a}(\theta_0) $, i.e.,
	$ {\bf w}_{k-1}={\bf a}(\theta_0) $,
	then for 
	$ \theta_k\neq\theta_0 $ and $ {\rho_k} $ satisfying 
	\begin{align}\label{eqn005}
	0\leq\rho_k\leq\min\left\{{\widetilde{\rho}_k},\overline{{\rho}}_k\right\},
	~\rho_k\neq{\widetilde{\rho}_k}
	\end{align}
	we have $ T({\beta}_{k,\star})={\mu}_{k,s}={\mu}_{k,\star} $.
	In other words, $ \textrm{C}^2\textrm{-WORD} $ obtains the same weight
	vector $ {\bf w}_{k} $ as ${\textrm A}^2{\textrm{RC}}$ in the above case.
\end{theorem2}
\begin{IEEEproof}
	See Appendix F.	
\end{IEEEproof}

Corollary 1 shows that $ \textrm{C}^2\textrm{-WORD} $
becomes equivalent to ${\textrm A}^2{\textrm{RC}}$
if $ {\bf w}_{k-1}={\bf a}(\theta_0) $ taken and the desired
level $ \rho_k $ satisfies specific condition, i.e., \eqref{eqn005}.
Note that the Corollary 1 has no 
limitation on the array configurations.
In addition, recalling {\it Remark 1}, we 
find that \eqref{eqn005} can be easily guaranteed.

In addition, another corollary of Proposition 4,
which relates to
centro-symmetric arrays \cite{centro1,centro2,centro3}, can be established.
In brief, a sensor array is called centro-symmetric if its element locations are symmetric with respect to the centroid and the complex characteristics of paired elements are the same.
To proceed,
we first define the conjugate centro-symmetric vector set $ \mathbb{V} $ as
\begin{align}
\mathbb{V}\triangleq\left\{{\bf v}\in\mathbb{C}^{N\times 1}\Big| {\bf v}(i)={\bf v}^{\ast}(N-i+1),~\forall i=1,\cdots,N\right\}.\nonumber
\end{align}	
It can be readily found that the steering vectors of a centro-symmetric array
locate in $ \mathbb{V} $, provided that the symmetric center is taken as
reference. Based on this observation, we can obtain
the second corollary
of Proposition 4.
\begin{theorem2}
	For a centro-symmetric array and the prescribed $ \theta_k $
	and $ \rho_k $, if $ {\bf w}_0\in{\mathbb{V}} $
	and
	\begin{align}
	\label{con1}&0\leq\rho_k\leq\min\left\{{\widetilde{\rho}_k},{\breve{\rho}_k}\right\},~\rho_k\neq{\widetilde{\rho}_k}\\
	\label{con2}&~~~~{\bf{w}}^{\mH}_{\bot}{\bf{a}}(\theta_0)
	{\bf{a}}^{\mH}(\theta_0){\bf{w}}_{\Arrowvert}\geq 0
	\end{align}
	hold true for $ k=1,2,\cdots $.
	Then $ \textrm{C}^2\textrm{-WORD} $ method will obtain the same
	weight vector as that of ${\textrm A}^2{\textrm{RC}}$ in every step of weight updating.
	In other words, these two approaches are completely equivalent in this scenario.
\end{theorem2}
\begin{IEEEproof}
	See Appendix G.
\end{IEEEproof}

The above Corollary 2 provides a case when
${\textrm A}^2{\textrm{RC}}$ becomes equivalent to $ \textrm{C}^2\textrm{-WORD} $. 
In this specific scenario, the parameter
selection of ${\textrm A}^2{\textrm{RC}}$ is optimal in the sense
of WNG.
For more general cases, the parameter determination of
${\textrm A}^2{\textrm{RC}}$ may not be the optimal one.
Therefore, the proposed $ \textrm{C}^2\textrm{-WORD} $ algorithm always
performs at least as good as ${\textrm A}^2{\textrm{RC}}$.

\textit{Remark 2}: Similar to {\it Remark 1}, in general we have ${\bf a}^{\mH}(\theta_k){\bf{a}}(\theta_k)>|{\bf a}^{\mH}(\theta_k){\bf{a}}(\theta_0)|$.
Based on this observation, the condition $ 0\leq{\rho_k}<{\widetilde{\rho}_k} $ can be
guaranteed as long as $ 0\leq{\rho_k}\leq1 $. Furthermore, to make $ 0\leq{\rho_k}\leq{\breve{\rho}_k} $
satisfied, we can simply restrict the desired response level $ {\rho_k} $ to be lower
than its previous level at $ \theta_k $, i.e.,
$ 0\leq{\rho_k}\leq{|{\bf{w}}^{\mH}_{k-1}{\bf a}(\theta_k)|}/
{|{\bf{w}}^{\mH}_{k-1}{\bf a}(\theta_0)|}=L_{k-1}(\theta_k,\theta_0) $.
Normally, $ L_{k-1}(\theta_k,\theta_0) $ is not greater than $ 1 $.
Therefore, \eqref{con1} can be easily satisfied as
long as $ 0\leq{\rho_k}\leq L_{k-1}(\theta_k,\theta_0) $.

\begin{algorithm}[!t]
	\caption{$ \textrm{C}^2\textrm{-WORD} $ based Pattern Synthesis Algorithm}\label{factorps}
	\begin{algorithmic}[1]
		\State give $ \theta_0 $, the desired pattern $ L_d(\theta) $,
		the initial weight vector $ {\bf w}_{0} $ and its corresponding  pattern $ L_0(\theta,\theta_0) $, set $ k=1 $
		\While{1}		
		\State select an angle $ \theta_k $ by comparing $ L_{k-1}(\theta,\theta_0) $ with
		$ ~~~~~L_d(\theta) $
		\State apply $ \textrm{C}^2\textrm{-WORD} $ to
		realize $L_k(\theta_k,\theta_0)=L_d(\theta_k) $, $ ~~~ $ $ ~~~~ $
		obtain $ {\bf w}_{k} $ in \eqref{e06} and the corresponding $ L_k(\theta,\theta_0) $
		\If{$ L_k(\theta,\theta_0){\rm ~is~not~satisfactory} $}
		\State set $ k=k+1 $
		\Else
		\State break
		\EndIf
		\EndWhile
		\State output $ {\bf w}_{k} $ and $ L_k(\theta,\theta_0) $		
	\end{algorithmic}
\end{algorithm}

\section{Pattern Synthesis via $ \textrm{C}^2\textrm{-WORD} $}
In this section, the application of $ \textrm{C}^2\textrm{-WORD} $ to pattern synthesis is briefly introduced.
Generally speaking, the strategy herein shares a similar concept of pattern synthesis
using ${\textrm A}^2{\textrm{RC}}$ \cite{refa2rc} or
WORD \cite{word}.
We synthesize a desirable beampattern by
successively adjusting the response levels at the directions where the specifications do not meet.

\begin{figure*}[!t]
	\centering
	\subfloat[]
	{\includegraphics[width=2.35in]{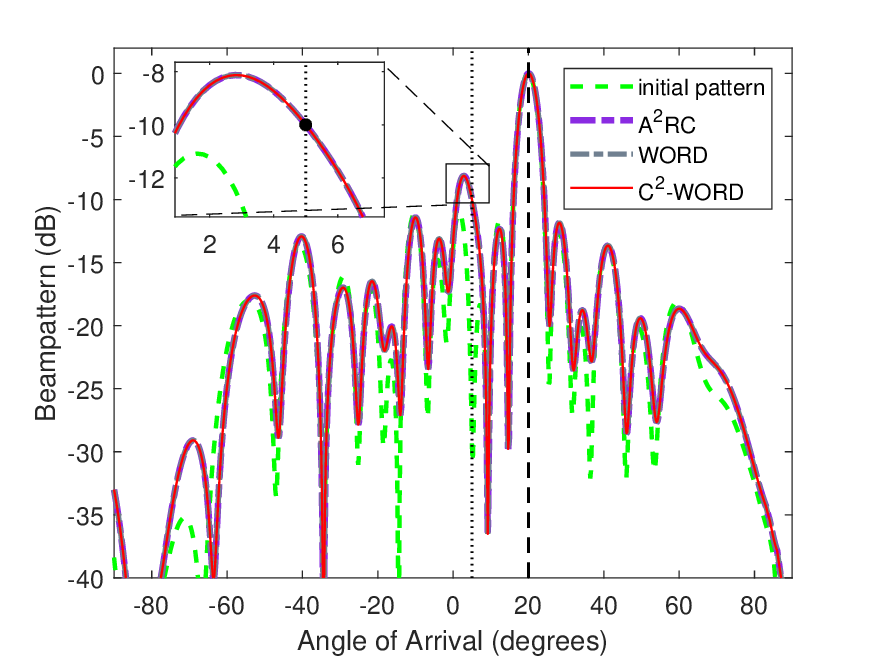}%
		\label{tspcontrolstep1allv3}}
	\hfil
	\subfloat[]
	{\includegraphics[width=2.35in]{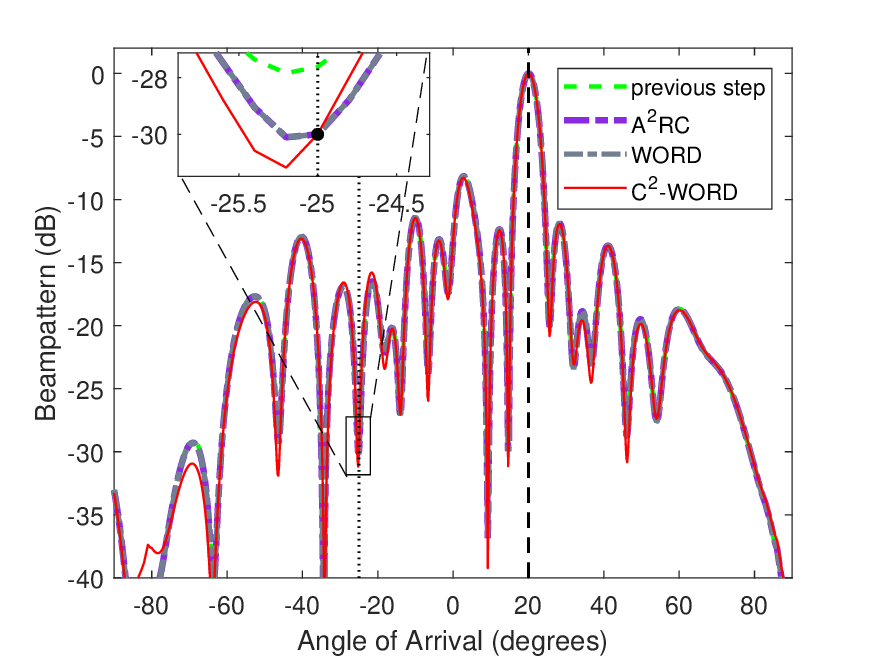}%
		\label{tspcontrolstep2allv3}}	
	\subfloat[]
	{\includegraphics[width=2.35in]{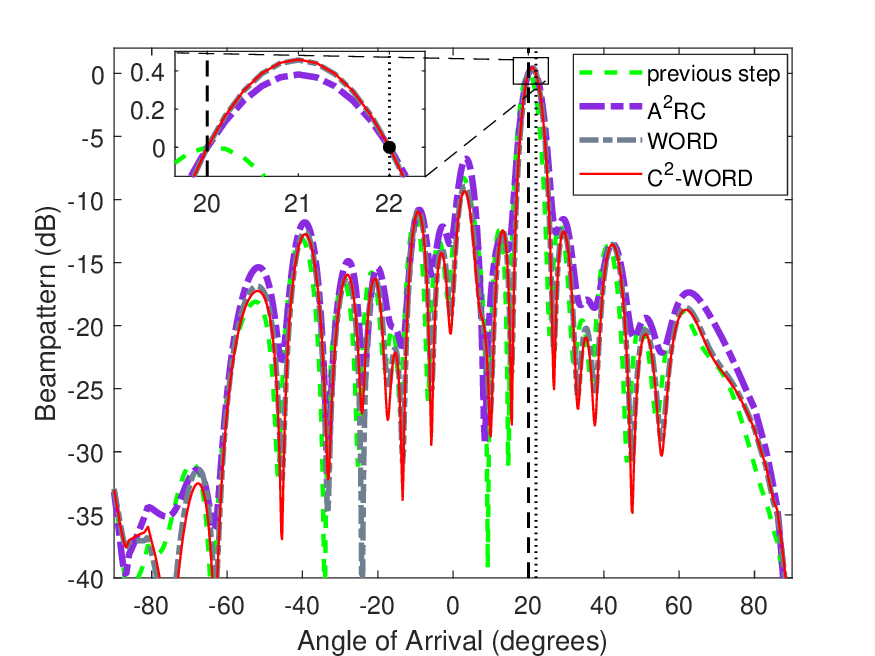}%
		\label{tspcontrolstep3allv3}}\\							
	\caption{Illustration of response control for a nonisotropic random array.
		(a) The synthesized beampatterns at the 1st step.
		(b) The synthesized beampatterns at the 2nd step.
		(c) The synthesized beampatterns at the 3rd step.}
	\label{tspcontrolst}
\end{figure*}

\begin{figure*}[!t]
	\centering
	\subfloat[]
	{\includegraphics[width=2.35in]{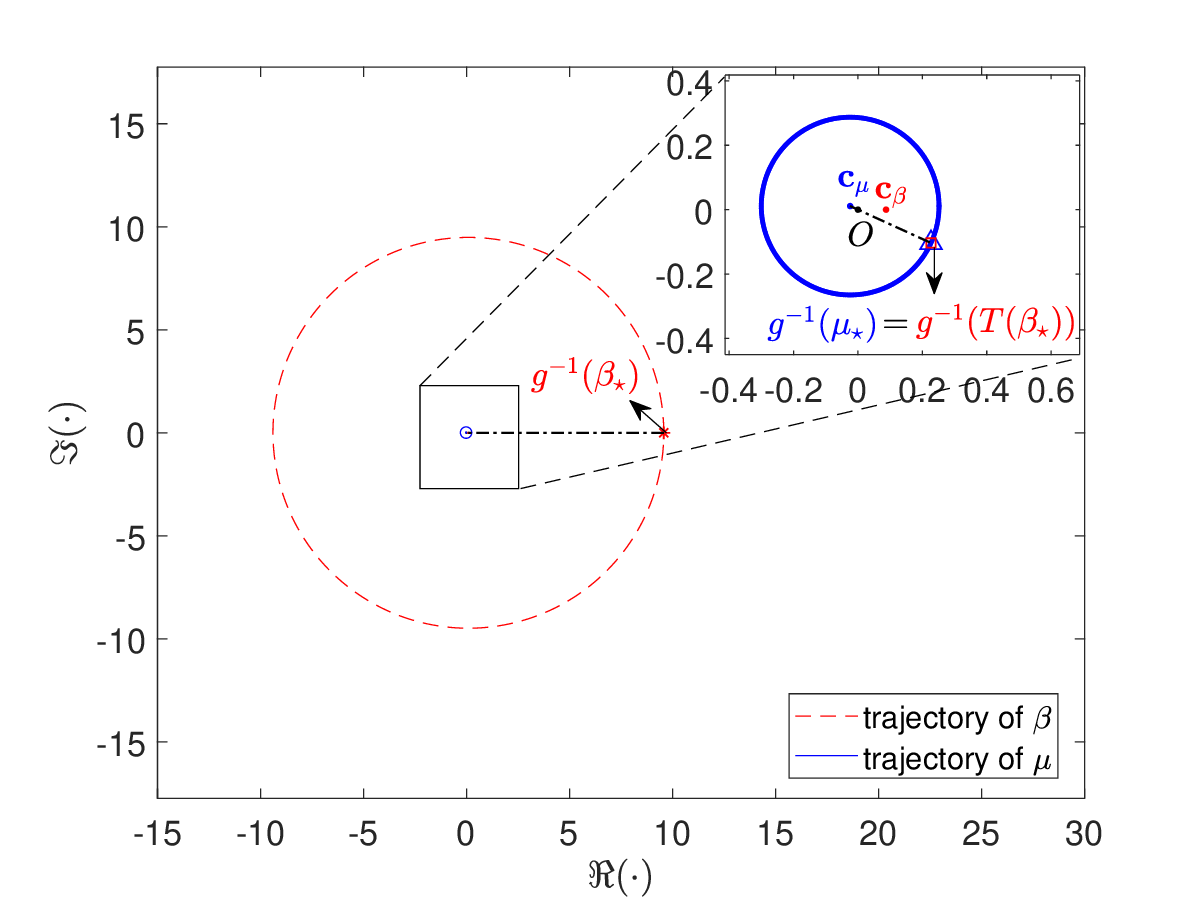}%
		\label{tspcirclebigstep1}}
	\hfil
	\subfloat[]
	{\includegraphics[width=2.35in]{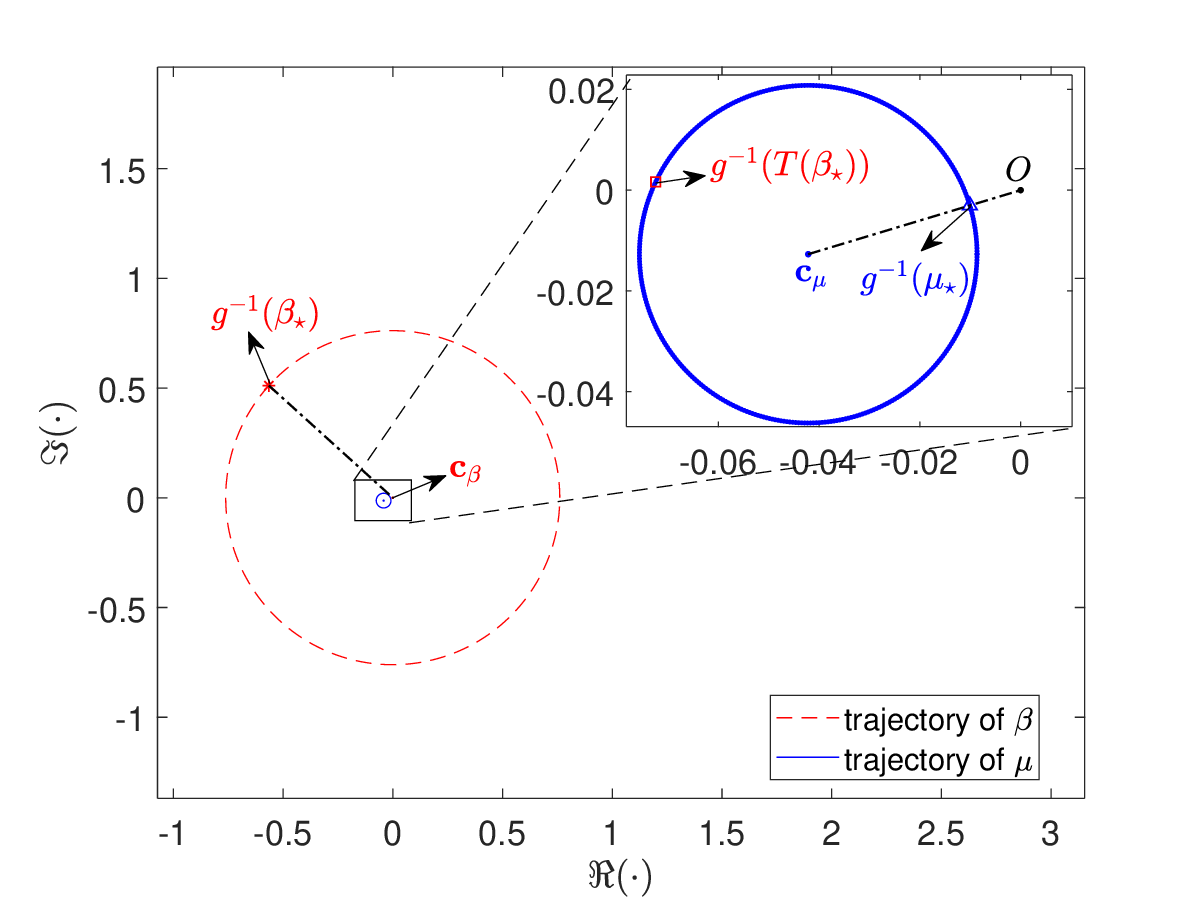}%
		\label{tspcirclebigstep2}}	
	\subfloat[]
	{\includegraphics[width=2.35in]{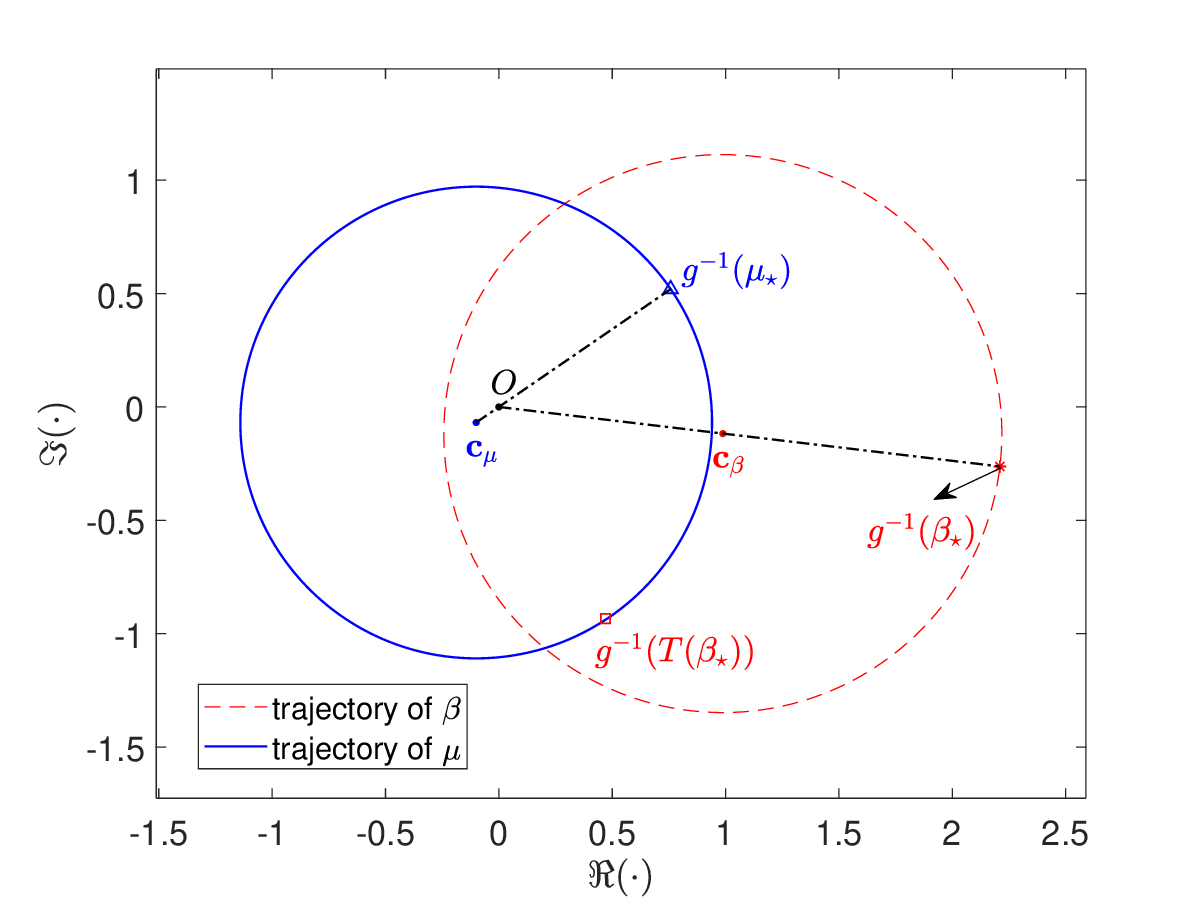}%
		\label{tspcirclebigstep3}}\\							
	\caption{Geometrical distributions of $ \beta $ and $ \mu $ at different steps.
		(a) Geometrical distributions of $ \beta $ and $ \mu $ at the 1st step.
		(b) Geometrical distributions of $ \beta $ and $ \mu $ at the 2nd step.
		(c) Geometrical distributions of $ \beta $ and $ \mu $ at the 3rd step.}
	\label{tspcontrolst3}
\end{figure*}

More precisely, let $ L_d(\theta) $ be the desired pattern. An initial pattern $ L_0(\theta,\theta_0) $ is firstly obtained by setting the weight vector
as $ {\bf w}_{0} $, and an angle $ \theta_1 $, at which the response level requires adjustment, is selected by comparing the initial pattern with the desired one. Next, the $ \textrm{C}^2\textrm{-WORD} $ scheme is applied to modify the weight vector $ {\bf w}_{0} $ to $ {\bf w}_{1} $, by setting the desired response level at $ \theta_1 $ as $ L_d(\theta_1) $. Similarly, by comparing $ L_{1}(\theta,\theta_0) $ with $ L_d(\theta) $, a second angle $ \theta_2 $, at which the response is needed to be adjusted, is selected. An updated weight vector $ {\bf w}_{2} $ can thus be achieved via $ \textrm{C}^2\textrm{-WORD} $. The above procedure is carried out successively once a satisfactory array pattern has been obtained.
As for how to select the angle to be controlled in every step, we follow
the strategy in \cite{refa2rc} and \cite{word}.
More specifically, for sidelobe synthesis, we select a peak angle
where the response difference (from the desired level) is relatively large.
For mainlobe synthesis, an angle
where the response deviates large from the desired one is chosen.
Finally, we summarize the $ \textrm{C}^2\textrm{-WORD} $ based pattern synthesis algorithm in Algorithm \ref{factorps}.

\section{Numerical Results}
In this section, simulations are presented to demonstrate $ \textrm{C}^2\textrm{-WORD} $ on array response control and array pattern synthesis.
For comparison purpose, the ${\textrm{A}}^2{\textrm{RC}}$ algorithm in \cite{refa2rc}, WORD algorithm in \cite{word},
convex programming (CP) method in \cite{ref24} and the semidefinite relaxation (SDR) method in \cite{ref27}
will also be tested if applicable.
Unless otherwise specified, 
we take $ {\bf a}(\theta_0) $ as the initial weight for
${\textrm{A}}^2{\textrm{RC}}$, WORD and the proposed $ \textrm{C}^2\textrm{-WORD} $.



\subsection{Response Control of a Nonisotropic Linear Random Array}
In this example, we illustrate the performance of $ \textrm{C}^2\textrm{-WORD} $
on array response control and show its advantage over ${\textrm A}^2{\textrm{RC}}$
and WORD.
More specifically, a 21-element nonisotropic linear random array (see e.g., \cite{ref070,ref07,refa2rc}) is considered. 
The pattern of the $ n $th element is given by
\begin{align}\label{gn}
\!\!\!g_n(\theta)=\left[{{\rm cos}
	\left( \pi l_n{\rm sin}(\theta+\zeta_n) \right)-{\rm cos}(\pi l_n)}\right]/
{{\rm cos}(\theta+\zeta_n)}
\end{align}
where $ \zeta_n $ and $ l_n $ represent the orientation and length of the element, respectively. More details of the array can be found in Table \ref{table7}, where the element positions (in wavelength) are
also specified.
To illustrate the effectiveness of the proposed method, 
we set the beam axis as $ \theta_0=20^{\circ} $ and pre-assign
the corresponding desired levels, see Table \ref{table10} for details.
The resulting WNGs of ${\textrm A}^2{\textrm{RC}}$, WORD and $ \textrm{C}^2\textrm{-WORD} $ are denoted by
$ G_{k,\rhd} $, $ G_{k,\times} $ and $ G_{k,\star} $, respectively.

\begin{table}[!t]
	\renewcommand{\arraystretch}{1.15}
	\caption{Parameters of the Nonisotropic Random Array}
	\label{table7}
	\centering
	\begin{tabular}{c | c | c | c | c | c | c | c}
		\hline
		$ n $&$ x_n(\lambda) $&$ l_n(\lambda) $&$\zeta_n(\rm{deg})$ & 
		$ n $&$ x_n(\lambda) $&$ l_n(\lambda) $&$\zeta_n(\rm{deg})$\\
		\hline
		1&0.00 &0.30  &0.0  & 12&5.50 &0.20  &5.0  \\
		2&0.45 &0.25 &-4.0  & 13&6.01 &0.29 &4.0  \\
		3&0.95 &0.24 &5.0  & 14&6.53 &0.20  &5.0  \\
		4&1.50 &0.20 &-32  & 15&7.07 &0.26 &-9.0  \\
		5&2.04 &0.26 &-3.2 & 16&7.52 &0.21 &7.0 \\
		6&2.64 &0.27 &10   & 17&8.00 &0.25 &10  \\
		7 &3.09 &0.23 &1.0 & 18&8.47 &0.21 &6.0  \\	
		8 &3.55 &0.24 &-10 & 19&8.98 &0.20  &-8.0  \\
		9 &4.05 &0.25 &0.0 & 20&9.53 &0.26 &0.0  \\
		10&4.55 &0.21 &7.0 & 21&10.01 &0.25 &5.0  \\
		11&5.06 &0.20  &5.0&   &     &     &    \\				
		\hline
	\end{tabular}
\end{table}
\begin{table}[!t]
	\renewcommand{\arraystretch}{1.15}
	\caption{Experiment Settings of Response Control and Obtained Parameters and WNGs of Different Methods}
	\label{table10}
	\centering
	\begin{tabular}{c | c | c | c}
		\hline
		& $ k=1 $ 			& $ k=2 $ 			& $ k=3 $\\
		\hline
		$ \theta_k $ & $ 5^{\circ} $ & $ -25^{\circ} $  & $ 22^{\circ} $ \\		
		\hline
		$ \rho_k $ & $ -10{~\rm dB} $ & $ -30{~\rm dB} $ & $ 0{~\rm dB} $\\	
		\hline
		$\mu_{k,\star}$      &0.2270-$ j $0.1034&-0.0101-$ j $0.0030  & 0.7577+$ j $0.5199 \\
		$G_{k,\rhd}$         &\textbf{12.9284}$ {~\rm dB} $           & 12.9336$ {~\rm dB} $ &12.1491$ {~\rm dB} $\\		
		\hline
		$\beta_{k,\times}$   &9.5716& 0.7606 &2.2132\\
		$G_{k,\times}$		 &\textbf{12.9284}$ {~\rm dB} $&12.9336$ {~\rm dB} $ &12.6085$ {~\rm dB} $\\
		$T(\beta_{k,\times})$ &0.2270-$ j $0.1034&-0.0101-$ j $0.0030 & 0.6499-$ j $0.7894 \\		
		\hline
		$\beta_{k,\star}$    &9.5716& -0.5649+$ j $0.5114  & 2.2097-$ j $0.2627 \\
		$G_{k,\star}$ &\textbf{12.9284}$ {~\rm dB} $&\textbf{12.9488}$ {~\rm dB} $ &\textbf{12.6184}$ {~\rm dB} $\\
		$T(\beta_{k,\star})$ &0.2270-$ j $0.1034&-0.0725+$ j $0.0016  & 0.4703-$ j $0.9336 \\		
		\hline
	\end{tabular}
\end{table}

\begin{figure*}[!t]
	\centering
	\subfloat[]
	{\includegraphics[width=2.35in]{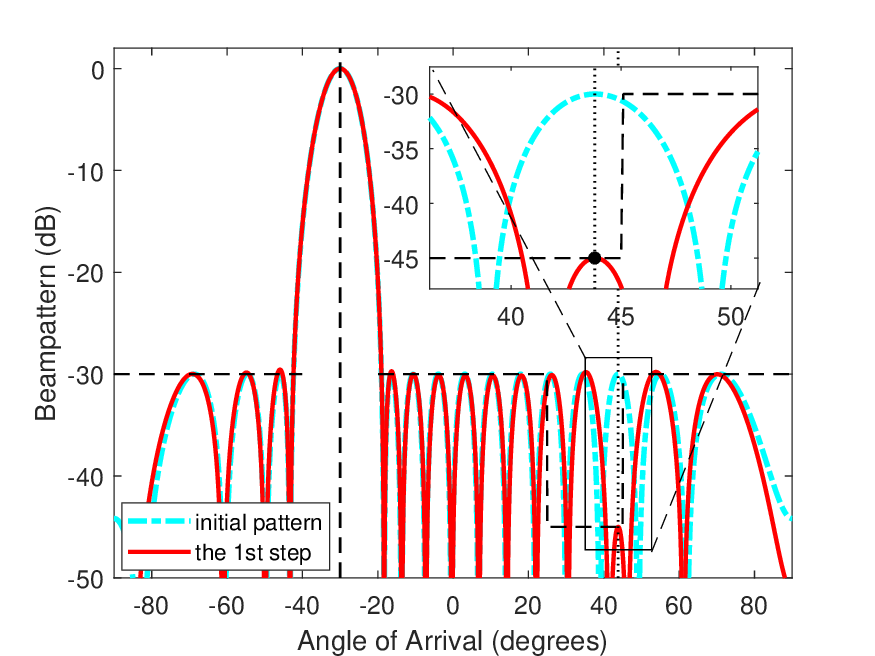}%
		\label{tapULAstep1}}
	\hfil
	\subfloat[]
	{\includegraphics[width=2.35in]{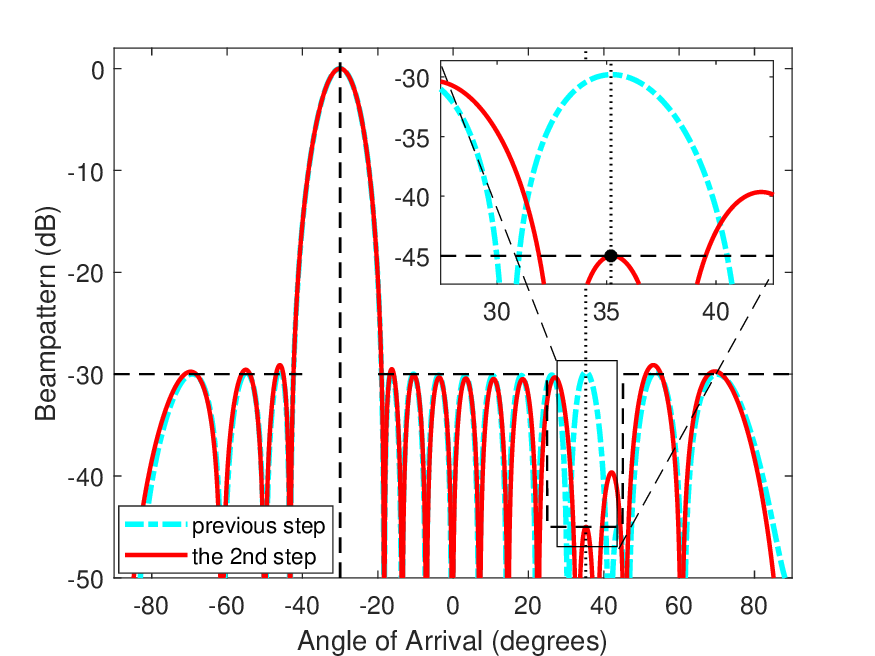}%
		\label{tapULAstep2}}	
	\subfloat[]
	{\includegraphics[width=2.35in]{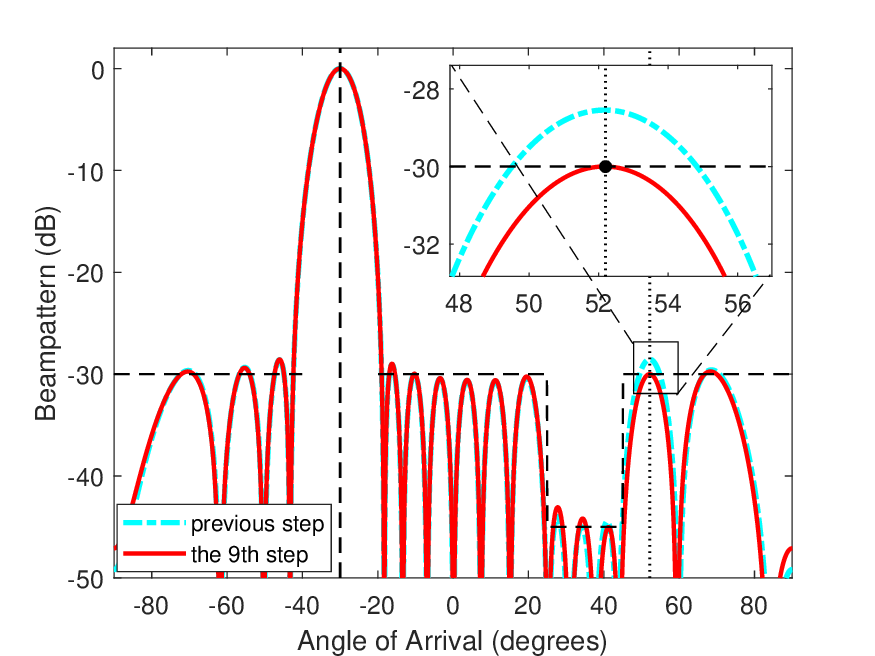}%
		\label{tapULAstep10}}\\							
	\caption{Synthesis procedure of nonuniform-sidelobe pattern using a ULA
		(starting from Chebyshev weight).
		(a) Synthesized pattern at the first step.
		(b) Synthesized pattern at the second step.
		(c) Synthesized pattern at the 9-th step.}
	\label{p2bf4}
\end{figure*}

Fig. \ref{tspcontrolst} depicts the obtained patterns at different steps.
Accordingly, Fig. \ref{tspcontrolst3} presents the trajectories of $ \beta $ and $ \mu $. 
It can be observed from Fig. \ref{tspcontrolst} that the responses can be adjusted as expected by
the three approaches. 
Table \ref{table10} summarizes the resulting parameters and WNGs.
One can see that the resulting WNG of $ \textrm{C}^2\textrm{-WORD} $ algorithm
is not less than those of ${\textrm A}^2{\textrm{RC}}$ and WORD for each response control step.
It is interesting to point out that $ T(\beta_{1,\star})=T(\beta_{1,\times})={\mu}_{1,\star} $,
which indicates that the three approaches obtain the same weight vectors
in the first step of response control.
An intuitive explanation of
this result is presented in Fig. \ref{tspcirclebigstep1}, where we can see that
the resulting $ \beta_{1,\star} $ is real and $ T(\beta_{1,\star}) $
coincides exactly with $ \mu_{1,\star} $.
This result is actually coincident with the conclusion of
Corollary 1. 
In the 2nd and 3rd steps, $ \textrm{C}^2\textrm{-WORD} $ obtains different beampatterns from those
of ${\textrm A}^2{\textrm{RC}}$ and WORD, see Fig. \ref{tspcontrolstep2allv3} and Fig. \ref{tspcontrolstep3allv3} for details.
The corresponding distributions of $ \beta $ and $ \mu $ are depicted in Fig. \ref{tspcirclebigstep2}
and Fig. \ref{tspcirclebigstep3}, from which we find that the resulting $ \beta_{k,\star} $'s are 
complex-valued and $ T(\beta_{k,\star}) $ are not
coincident with $ \mu_{k,\star} $, $ k=2,3 $.

\subsection{Pattern Synthesis Using $ \textrm{C}^2\textrm{-WORD} $}
In this section, representative simulations are presented to
illustrate the application of 
$ \textrm{C}^2\textrm{-WORD} $ to pattern synthesis.
\subsubsection{Nonuniform Sidelobe Synthesis for a ULA} In the first example, a 16-element uniformly
spaced linear array (ULA) is considered. We steer the beam axis to $ \theta_0=-30^{\circ} $.
The desired beampattern has nonuniform sidelobe levels. More specifically, the upper level
is $ -45{\rm dB} $ in the region $ [25^{\circ},45^{\circ}] $ and 
$ -30{\rm dB} $ in the rest of the sidelobe region.
To lower the complexity,
we take the initial weight of $ \textrm{C}^2\textrm{-WORD} $ as the Chebyshev weight with a $ -30{\rm dB} $
sidelobe attenuation. Note that the initial weight is a conjugate centro-symmetric vector, i.e., $ {\bf w}_0\in{\mathbb{V}} $.

\begin{figure}[!t]
	\centering
	\includegraphics[width=3.0in]{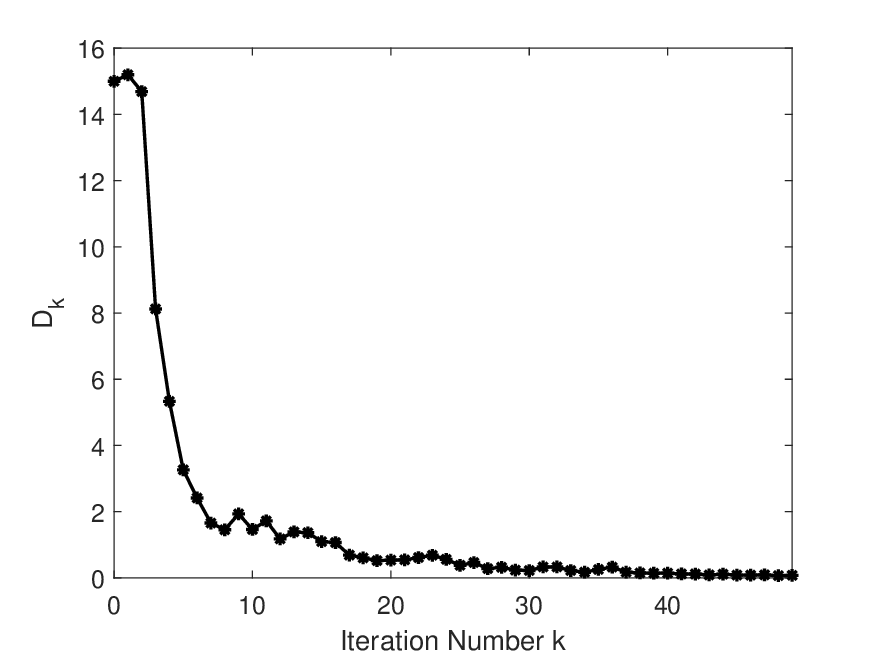}
	\caption{Maximum response deviation $ D_k $ versus the iteration number.}
	\label{delta}
\end{figure}

Fig. \ref{p2bf4} presents several intermediate results when synthesizing pattern by $ \textrm{C}^2\textrm{-WORD} $.
At each response control step, we select one sidelobe peak and then adjust the response to
its desired level. From Fig. \ref{p2bf4}, one can see that the resulting response envelope
is similar to the desired one after conducting 9 response control steps.
To explore the convergence of the proposed approach, we define $ D_k $ as the maximum response deviation within the set of the sidelobe peak angles at the $k$th step (denoted as $ {\Omega}^{k}_s $), i.e.,
\begin{align}
D_k\triangleq
\max_{\theta\in{\Omega}_s^{k}}
\left(
L_{k}(\theta,\theta_0)-L_d(\theta)
\right).
\end{align}	
The curve of $ D_k $ versus the iterative number $ k $ is depicted in
Fig. \ref{delta}, which clearly shows that $ D_k $ decreases
with the increase of iteration.
After carrying out 40 response control steps, the resulting $ D_k $ 
equals approximately to zero and we terminate the synthesis process.

\begin{figure}[!t]
	\centering
	\includegraphics[width=3.0in]{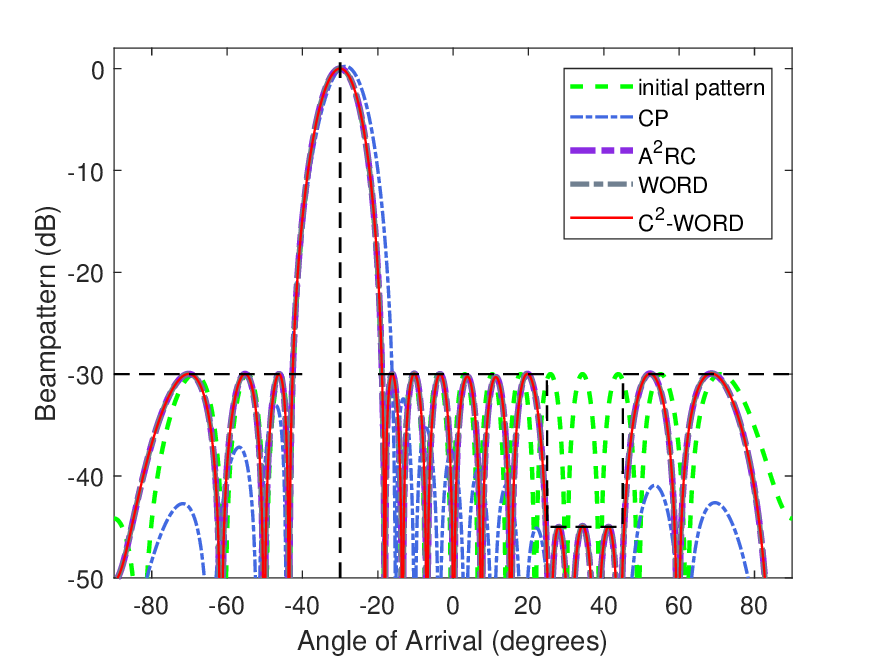}
	\caption{Comparison of the synthesized patterns.}
	\label{final}
\end{figure}

The ultimate pattern of the proposed approach is presented in
Fig. \ref{final}, where the results of CP, ${\textrm A}^2{\textrm{RC}}$ and WORD are
also displayed. For both ${\textrm A}^2{\textrm{RC}}$ and WORD, we carry out the same iteration steps 
as that of $ \textrm{C}^2\textrm{-WORD} $, and the resulting beampatterns of
these three approaches are identical in this case.
This result is consistent with the prediction of Corollary 2.
In addition, we can see that the resulting mainlobe width of CP method is wider,
although a qualified sidelobe level is obtained.

\begin{figure}[!t]
	\centering
	\includegraphics[width=3.0in]{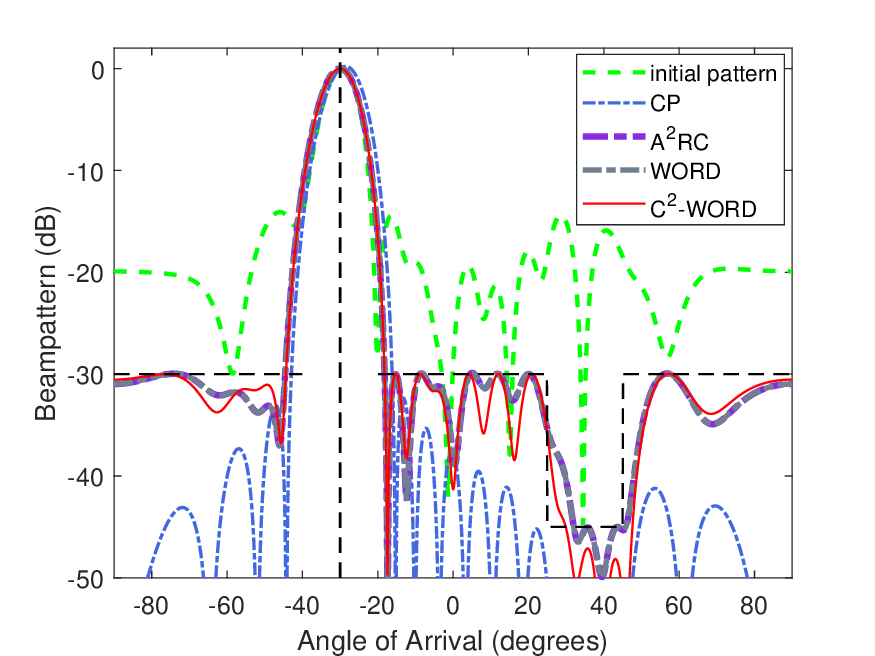}
	\caption{Synthesized patterns with non-uniform sidelobe for a ULA with mutual coupling.}
	\label{nonuniformC}
\end{figure}

To further examine the performance of the proposed approach.
We take the mutual coupling into consideration with other settings unchanged.
The channel isolation \cite{add5} between different elements is $ -20{\rm dB} $.
The resulting beampatterns of different approaches are presented in Fig. \ref{nonuniformC}.
In this case, the proposed $ \textrm{C}^2\textrm{-WORD} $ algorithm obtains
a different pattern from those of ${\textrm A}^2{\textrm{RC}}$ and WORD.
The resulting WNG of our algorithm is $ {11.30 \rm dB} $, which is
higher than the corresponding $ {11.01 \rm dB} $ of ${\textrm A}^2{\textrm{RC}}$ and WORD.
In addition, the resulting WNG of CP method is $ {10.78 \rm dB} $ and
the synthesized beampattern is not aligned with the desired one.

\begin{table}[!t]
	\centering
	\renewcommand{\arraystretch}{1.15}
	{\caption{Element Locations of Linear Random Array and Weights Obtained by the Proposed Method}
		\label{tabdl1}		
		\begin{tabular}{c | c| c || c | c| c }
			\hline
			{$n$}&$ x_n(\lambda) $& {$ ~~w_n~~ $} &{$n$}&$ x_n(\lambda) $& {$ ~~w_n~~ $}\\
			\hline
			{1}&{0.00}&{$ 0.2523e^{+j0.0468} $}  &{9} &{3.99}& {$ 0.8648e^{-j2.8313} $}\\
			{2}&{0.47}&{$ 0.0239e^{+j0.6417} $}  &{10}&{4.48} & {$ 0.0802e^{+j1.0755} $}\\
			{3}&{1.01}&{$ 0.1932e^{-j2.3044} $}  &{11}&{4.96} & {$ 0.9415e^{+j0.2068} $}\\
			{4}&{1.47}&{$ 0.3330e^{+j3.0465} $}  &{12}&{5.43} & {$ 1.0000e^{+j0.0084} $}\\
			{5}&{1.97}&{$ 0.2863e^{+j0.3971} $}  &{13}&{5.94} & {$ 0.7295e^{+j0.0597} $}\\
			{6}&{2.54}&{$ 0.6463e^{+j0.2904} $}  &{14}&{6.49} & {$ 0.7338e^{+j0.0040} $}\\
			{7}&{3.06}&{$ 0.1202e^{+j0.0574} $}  &{15}&{6.98} & {$ 0.6187e^{+j0.1650} $}\\
			{8}&{3.53}&{$ 0.7930e^{-j2.8431} $}  &{16}&{7.46} & {$ 0.3340e^{-j0.2297} $}\\						
			\hline
	\end{tabular}}
\end{table}

\begin{figure}[!t]
	\centering
	\includegraphics[width=3.0in]{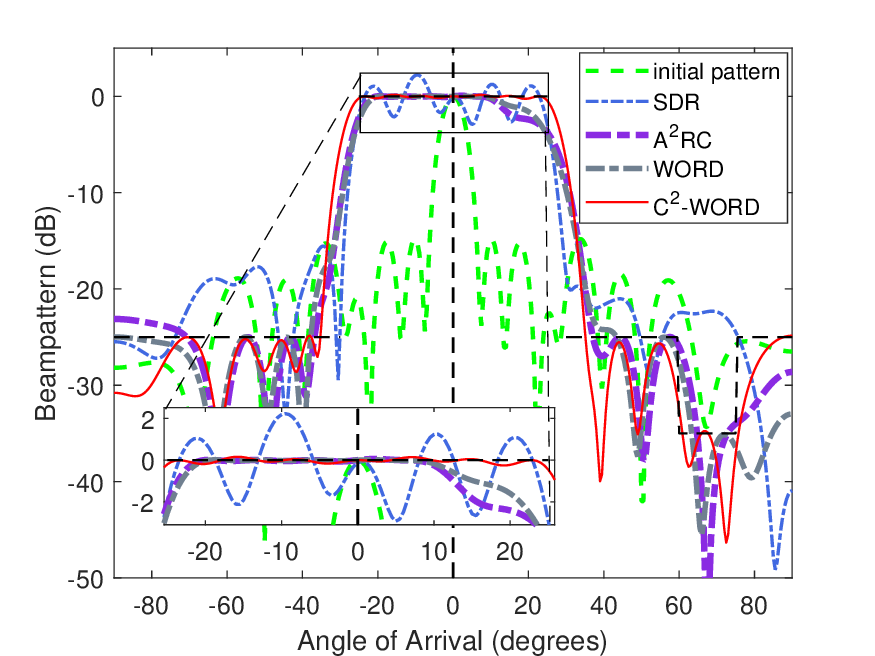}
	\caption{Synthesized patterns with flat-top mainlobe and broad-notch sidelobe for a
		random linear array with mutual coupling.}
	\label{topfigok}
\end{figure}

\subsubsection{Synthesizing Flat-Top Pattern for Linear Random Array}
In this example, we consider a linear random array, with its element location specifying in Table \ref{tabdl1}.
In this case, the desired pattern has both sidelobe and mainlobe constraints.
More specifically, the upper level is
$ -35{\rm dB} $ in the region $ [60^{\circ},75^{\circ}] $ and 
$ -25{\rm dB} $ in the rest of the sidelobe region.
For mainlobe response,
its desired level is $ 0{\rm dB} $ in the pre-assigned region $ [-25^{\circ},25^{\circ}] $.
In addition, we consider mutual coupling with a $ -25{\rm db} $ channel isolation between different elements.
After carrying out 300 response control steps,
the proposed $ \textrm{C}^2\textrm{-WORD} $ approach synthesizes a desirable beampattern and
the resulting weightings are listed in Table \ref{tabdl1}.

Fig. \ref{topfigok} presents the resulting patterns of $ \textrm{C}^2\textrm{-WORD} $,
${\textrm A}^2{\textrm{RC}}$, WORD (all with the same iteration steps) and SDR.
We can see that the $ \textrm{C}^2\textrm{-WORD} $ approach obtains satisfactory 
responses at both sidelobe and mainlobe region.
The ripple of the mainlobe response is about $ 0.3{\rm dB} $,
which is less than those of the other three methods.
Note that both ${\textrm A}^2{\textrm{RC}}$ and WORD lead to
some pattern distortions at the mainlobe region.
For the SDR approach, the resulting mainlobe ripple is large and the synthesized
sidelobe is not qualified. The reason probably is that the ultimate weight
of SDR method may not satisfy the pre-defined constraints, due to the relaxation operator
to the original problem.

\begin{figure}[!t]
	\centering
	\includegraphics[width=3.0in]{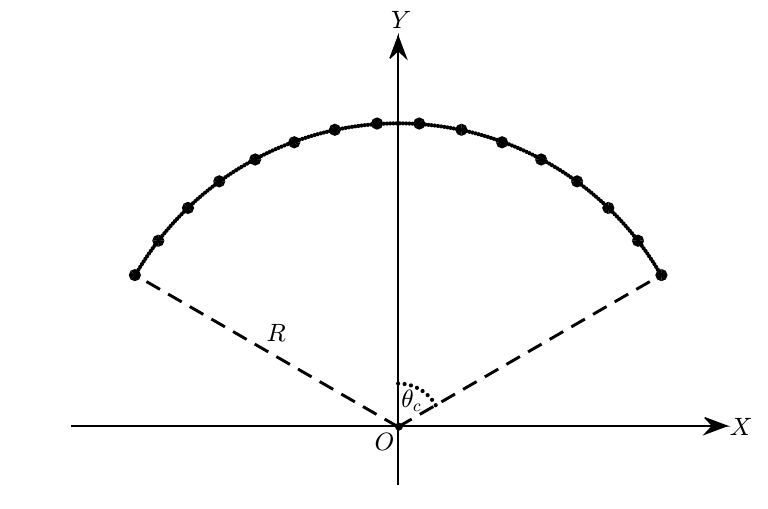}
	\caption{Illustration of a circular arc array.}
	\label{Fig8}
\end{figure}

\begin{figure}[!t]
	\centering
	\includegraphics[width=3.0in]{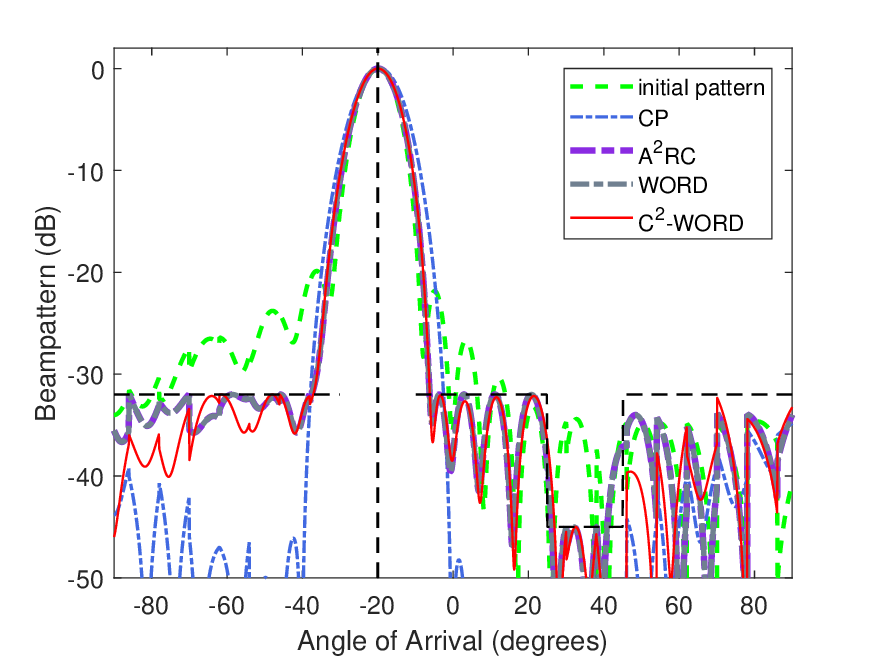}
	\caption{Synthesized patterns for a circular arc array.}
	\label{nonistropicfigok}
\end{figure}

\subsubsection{Pattern Synthesis for Conformal Array}
To further show the effectiveness of the proposed $ \textrm{C}^2\textrm{-WORD} $ algorithm,
we follow the array configuration in \cite{2017-2} and consider
a circular arc array that conforms to a cylindrical surface as shown in Fig. \ref{Fig8}.
The element number is $ 16 $ and the distance between adjacent elements is
half a wavelength. The $ \theta_c $ in Fig. \ref{Fig8} is set as $ 60^{\circ} $.
In addition, we take the element polarized pattern and mutual coupling effect into consideration.
More specifically, the element pattern follows the lowest order circular patch model
\cite{add11,add21,add3}, and the beam pattern can be analytically expressed as 
described in \cite{add4}. 
As for mutual coupling, the corresponding channel isolation between different elements is $ -25{\rm db} $.
In this case, we steer the beam axis to $ \theta_0=-20^{\circ} $.
The desired pattern has $ -45{\rm dB} $ upper level in the region $ [25^{\circ},45^{\circ}] $ and 
is expected to be lower than
$ -32{\rm dB} $ in the rest of the sidelobe region.
For simplicity, we only consider the pattern that is coplanar to the array plane, although
the extensions are straightforward.
Fig. \ref{nonistropicfigok} depicts the resulting patterns of different approaches.
One can see that our approach 
synthesizes a more satisfactory pattern than
those of CP, ${\textrm A}^2{\textrm{RC}}$ and WORD.


\section{Conclusions}
In this paper, we have presented a new scheme named
complex-coefficient weight vector orthogonal decomposition
($ \textrm{C}^2\textrm{-WORD} $).
The $ \textrm{C}^2\textrm{-WORD} $ algorithm is modified from the existing WORD method.
We have extended the WORD algorithm by allowing a complex-valued combining coefficient
in the devised $ \textrm{C}^2\textrm{-WORD} $ algorithm.
Moreover, parameter selection has been carried out to
maximize the white noise gain (WNG), and a closed-form solution
of weight vector updating has been obtained.
In addition, we have presented 
the benefits of $ \textrm{C}^2\textrm{-WORD} $ (comparing to WORD),
and have discussed the connection between $ \textrm{C}^2\textrm{-WORD} $
and the existing ${\textrm A}^2{\textrm{RC}}$ algorithm.
It has been shown that $ \textrm{C}^2\textrm{-WORD} $ may degrade into
${\textrm A}^2{\textrm{RC}}$ under specific circumstances,
and alway performs at least as good as
${\textrm A}^2{\textrm{RC}}$ and WORD in the sense of WNG.
The application of $ \textrm{C}^2\textrm{-WORD} $
to array pattern synthesis has been studied
and validated with various examples.
Based on the fundamentals developed in this paper,
a further application of $ \textrm{C}^2\textrm{-WORD} $ to robust
sidelobe control and synthesis will be considered in \cite{robust},
by taking the array steering vector uncertainties
into consideration.


%

\appendices
\section{Proof of Lemma 1}
For simplicity, we omit the subscript $ k $ of $ \beta $ and
$ \widetilde{\beta} $ in sequel.
To proof Lemma 1, we can find a specific $ \widetilde{\beta}\in\mathbb{C} $ and
the corresponding $ \widetilde{{\bf w}}_{k} $ in \eqref{eqn1023} making \eqref{eq12} satisfied.
To this end, for any given $ \beta\in\mathbb{R} $, we set $ \widetilde{\beta} $ as
\begin{align}\label{eqn1122}
\widetilde{\beta}=\beta+j\zeta
\end{align}
where $ \zeta $ is defined as
\begin{align}\label{eqn1123}
\zeta\triangleq{2{\beta}^2{{\eta}_i}}\big/({|{\xi}_{\bot}|^2+2{\beta}{\eta}_r})\in\mathbb{R}
\end{align}
with $ {\eta}_r={\Re}({\xi}_{\Arrowvert}{\xi}^{*}_{\bot}) $, 
$ {\eta}_i={\Im}({\xi}_{\Arrowvert}{\xi}^{*}_{\bot}) $,
$ {\xi}_{\Arrowvert}={\bf{w}}^{\mH}_{\Arrowvert}{\bf a}(\theta_0) $, 
$ {\xi}_{\bot}={\bf{w}}^{\mH}_{\bot}{\bf a}(\theta_0) $.
Since both $ \beta $ and $ \zeta $ are real-valued, we have
\begin{align}\label{eqn118}
{\Re}(({\beta}-j{\zeta}){\xi}_{\Arrowvert}{\xi}^{*}_{\bot})=
{\beta}{{\eta}_{r}}+\zeta{{\eta}_i}.
\end{align}

On the other hand, one can readily obtain from \eqref{eqn1123} that
$ 2{\beta}^2{{\eta}_i}=\zeta(|{\xi}_{\bot}|^2+2{\beta}{\eta}_r) $.
On this basis, it can be easily derived that 
$ 2{\beta}^2\zeta{{\eta}_i}+2{\beta}^3{{\eta}_{r}}={\zeta}^2(|{\xi}_{\bot}|^2+2{\beta}{\eta}_r)+
2{\beta}^3{{\eta}_{r}} $,
which can be equivalently re-expressed as
\begin{align}\label{eqn1126}
2{\beta}^2({\beta}{{\eta}_{r}}+\zeta{{\eta}_i})={\zeta}^2|{\xi}_{\bot}|^2+2({\beta}^2+{\zeta}^2)
{\beta}{{\eta}_{r}}.
\end{align}

Combining \eqref{eqn118} and \eqref{eqn1126}, one can obtain that
\begin{align}\label{eqn1127}
&{\beta}^2
\underbrace{\left(
	|{\xi}_{\bot}|^2+({\beta}^2+{\zeta}^2)|{\xi}_{\Arrowvert}|^2+
	2{\Re}(({\beta}-j{\zeta}){\xi}_{\Arrowvert}{\xi}^{*}_{\bot})
	\right)}_{|{\xi}_{\bot}+({\beta}-j{\zeta}){\xi}_{\Arrowvert}|^2}\nonumber\\
&~~~~~~~~~~~~~~~~~
=({\beta}^2+{\zeta}^2)
\underbrace{
	\left(
	|{\xi}_{\bot}|^2+{\beta}^2|{\xi}_{\Arrowvert}|^2+2{\beta}{\eta_r}
	\right)}_{|{\xi}_{\bot}+{\beta}{\xi}_{\Arrowvert}|^2}
\end{align}
and then
\begin{align}\label{eqn1137}
\dfrac{{({\beta}^2+{\zeta}^2)}}
{{|{\xi}_{\bot}+({\beta}-j{\zeta}){\xi}_{\Arrowvert}|^2}}=
\dfrac{{{\beta}^2}}{|{\xi}_{\bot}+{\beta}{\xi}_{\Arrowvert}|^2}.
\end{align}
Substituting $ \widetilde{{\bf w}}_{k}={\bf{w}}_{\bot}+\widetilde{\beta}{\bf{w}}_{\Arrowvert} $ into
$ L_{k}(\theta_k,\theta_0) $ and combining \eqref{eqn1137}, we have
\begin{align}
L_{k}(\theta_k,\theta_0)\big|_{{\bf w}=\widetilde{{\bf w}}_{k}}\!\!&=
\dfrac{{({\beta}^2+{\zeta}^2)|{\bf{w}}^{\mH}_{\Arrowvert}{\bf a}(\theta_k)|^2}}{{|{\xi}_{\bot}+({\beta}-j{\zeta}){\xi}_{\Arrowvert}|^2}}
\nonumber\\
&=\dfrac{{{\beta}^2|{\bf{w}}^{\mH}_{\Arrowvert}{\bf a}(\theta_k)|^2}}
{{|{\xi}_{\bot}+{\beta}{\xi}_{\Arrowvert}|^2}}\nonumber\\
&=L_{k}(\theta_k,\theta_0)\big|_{{\bf w}={{\bf w}}_{k}}
\end{align}
which indicates that \eqref{eq12} holds true.
Moreover, we know from \eqref{eqn1122}
that $ \widetilde{\beta}\notin\mathbb{R} $ if only $ \beta\neq 0 $ and
$ \eta_i\neq 0 $. This completes the proof of Lemma 1.

\section{Proof of Proposition 1}
For the sake of clarity, we may omit the subscript $ k $ in sequel.
To solve $ {\bf{z}}^{\mH}{\bf{B}}{\bf{z}}=0 $, we take eigenvalue decomposition of $ {\bf{B}} $, i.e.,
$ {\bf{B}}={{\bf U}}{\bf\Lambda}{{\bf U}}^{\mH} $,
where $ {\bf U} $ is an unitary matrix, $ {\bf\Lambda}={\rm Diag}([\lambda_1,\lambda_2]) $ with
$ \lambda_1 $ and $ \lambda_2 $ denoting eigenvalues of $ {\bf B} $.
Define $ {\bf{y}}\triangleq{{\bf U}}^{\mH}{\bf{z}} $.
Then $ {\bf{z}}^{\mH}{\bf{B}}{\bf{z}}=0 $ can be equivalently expressed as
$ {\bf{y}}^{\mH}{\bf\Lambda}{\bf{y}}=0 $, and further
\begin{align}\label{tap01211}
{{\lambda}_1}|{\bf{y}}(1)|^2+{{\lambda}_2}|{\bf{y}}(2)|^2=0.
\end{align}
Since $ {{\lambda}_1}{{\lambda}_2}=\mbox{det}({\bf{B}})=-\rho_k| {\bf{w}}^{\mH}_{\bot}{\bf{a}}(\theta_0) |^2
| {\bf{w}}^{\mH}_{\Arrowvert}{\bf{a}}(\theta_k) |^2\leq 0 $,
we learn that \eqref{tap01211} can be solved.
Let us first denote that
\begin{align}\label{f016}
{\bf U}=\begin{bmatrix}u_{11}&u_{12}  \\ u_{21}&u_{22} \end{bmatrix}.
\end{align}
Since $ {\bf{y}}={{\bf U}}^{\mH} {\bf{z}}$, we have $ {\bf{y}}(1)=u^{\ast}_{11}+u^{\ast}_{21}{\beta} $
and $ {\bf{y}}(2)=u^{\ast}_{12}+u^{\ast}_{22}{\beta} $.
Then, one can obtain that
\begin{align}\label{tap0024r1}
&{{\lambda}_1}|{\bf{y}}(1)|^2+{{\lambda}_2}|{\bf{y}}(2)|^2\nonumber\\
&=
{{\lambda}_1}|u^{\ast}_{11}+u^{\ast}_{21}{\beta}|^2+
{{\lambda}_2}|u^{\ast}_{12}+u^{\ast}_{22}{\beta}|^2\nonumber\\
&=\underbrace{{{\lambda}_1}|u_{11}|^2+{{\lambda}_2}|u_{12}|^2}_{{\bf B}(1,1)}+
\underbrace{({{\lambda}_1}u^{\ast}_{21}u_{11}+{{\lambda}_2}u^{\ast}_{22}u_{12})}_{{\bf B}(1,2)}{\beta}+\nonumber\\
&~~~~~~\underbrace{({{\lambda}_1}u^{\ast}_{11}u_{21}+{{\lambda}_2}u^{\ast}_{12}u_{22})}_{{\bf B}(2,1)}{\beta}^{\ast}
+\underbrace{({{\lambda}_1}|u_{21}|^2+{{\lambda}_2}|u_{22}|^2)}_{{\bf B}(2,2)}|{\beta}|^2\nonumber\\
&={\bf B}(1,1)+2{\Re}\left({\bf B}(1,2)\cdot{\beta}\right)+{\bf B}(2,2)|{\beta}|^2\nonumber\\
&={\bf B}(1,1)+2{\Re}\left({\bf B}(1,2)\right)\cdot{\Re}({\beta})-2{\Im}\left({\bf B}(1,2)\right)\cdot{\Im}({\beta})\nonumber\\
&~~~~~~~+{\bf B}(2,2)\left({\Re}^2({\beta})+{\Im}^2({\beta})\right)\nonumber\\
&=0
\end{align}
where we have utilized the identities that
\begin{subequations}
\begin{align}
{\bf {B}}(1,1)&={{\lambda}_1}|u_{11}|^2+{{\lambda}_2}|u_{12}|^2\\
{\bf {B}}(1,2)&={{\lambda}_1}u^{\ast}_{21}u_{11}+{{\lambda}_2}u^{\ast}_{22}u_{12}\\
{\bf {B}}(2,1)&={{\lambda}_1}u^{\ast}_{11}u_{21}+{{\lambda}_2}u^{\ast}_{12}u_{22}\\
{\bf {B}}(2,2)&={{\lambda}_1}|u_{21}|^2+{{\lambda}_2}|u_{22}|^2.
\end{align}
\end{subequations}
Eqn. \eqref{tap0024r1} can be rewritten as
\begin{align}\label{f005}
&\left(
{\Re}({\beta})+\frac{{\Re}\left({\bf {B}}(1,2)\right)}{{\bf {B}}(2,2)}
\right)^2+
\left(
{\Im}({\beta})-\frac{{\Im}\left({\bf {B}}(1,2)\right)}{{\bf {B}}(2,2)}
\right)^2\nonumber\\
&~~~~=-\dfrac{{\bf {B}}(1,1)}{{\bf {B}}(2,2)}+\dfrac{|{\bf {B}}(1,2)|^2}{{\bf {B}}^2(2,2)}
=-\dfrac{{\rm det}({\bf {B}})}{{\bf {B}}^2(2,2)}
\end{align}
which implies that $ \left[{\Re}({\beta})~{\Im}({\beta})\right]^{\mT} $
locates in a circle with center 
$ {\bf c}_{\beta} = 
\begin{bmatrix}-{\Re}\left({\bf B}(1,2)\right)& {\Im}\left({\bf B}(1,2)\right)\end{bmatrix}^{\mT}/{\bf B}(2,2) $
and radius $ R_{\beta}={\sqrt{-{\rm det}({\bf B})}}/{|{\bf B}(2,2)|} $.
This completes the proof.

\section{Proof of Proposition 2}
Substituting \eqref{criterion4} into 
\eqref{mini_obj} and recalling the constraint \eqref{word005}, we have
\begin{align}
G_k \propto J(\beta_k)\triangleq
\dfrac{|\beta_k|^2|{\bf{w}}^{\mH}_{\Arrowvert}{\bf a}(\theta_k)|^2}
{\|{\bf{w}}_{\bot}\|^2_2+|\beta_k|^2\|{\bf{w}}_{\Arrowvert}\|^2_2}.
\end{align}
%
Moreover, it is not hard to derive that
\begin{align}\label{e2067}
\dfrac{\partial{J(\beta_k)}}
{\partial{|\beta_k|^2}}=
\dfrac{|{\bf{w}}^{\mH}_{\Arrowvert}{\bf a}(\theta_k)|^2\|{\bf{w}}_{\bot}\|^2_2}
{\left({\|{\bf{w}}_{\bot}\|^2_2+|\beta|^2\|{\bf{w}}_{\Arrowvert}\|^2_2}\right)^2}\geq 0
\end{align}
which implies that the optimization function $ J(\beta_k) $ is monotonically non-decreasing
with the increase of $ |\beta_k| $.
According to this observation, one can readily obtain that
\begin{align}\label{e2069}
{\beta_{k,\star}}={\rm arg}~\max_{{\beta_k}\in{\mathbb{C}_{{\beta}_k}}}~|\beta_k|.
\end{align}

Recalling Proposition 1 and Fig. \ref{tspcirclebetabasic}, we know that
the $ \beta_k $ with maximum modulo among $ {\mathbb{C}_{{\beta}_k}} $
is the intersection point of circle $ \mathbb{C}_{\beta_k} $ with the line that passes $ {O} $ and $ {\bf c}_{\beta_k} $, see $ g^{-1}({\beta}_{k,l}) $ in 
Fig. \ref{tspcirclebetabasic}.
Mathematically, it can be readily obtained that
\begin{align}\label{beta_opt}
{\beta_{k,\star}}=(|{\bf c}_{\beta_k}|+{R_{\beta_k}})e^{j\angle g({\bf c}_{\beta_k})}.
\end{align}
This completes the proof.

\section{Proof of Proposition 3}
We divide the proof details of Proposition 3 into the following 3 steps.

{\it Step 1:} In the first step, we will show that the optimal solution (denoted as $ {\bf w}_{\triangleright} $) of problem \eqref{eqn011} satisfies
\begin{align}\label{key005}
{\bf w}_{\triangleright}\in\mathcal{R}([{\bf a}(\theta_0),{\bf a}(\theta_k)]).
\end{align}

To this end, it should be noted that the space $ {\mathbb{C}}^{N} $ satisfies
\begin{align}
{\mathbb{C}}^{N}=\mathcal{R}([{\bf a}(\theta_0),{\bf a}(\theta_k)])\oplus
\mathcal{R}^{\bot}([{\bf a}(\theta_0),{\bf a}(\theta_k)]).
\end{align} 
In other words, we can split $ \forall{\bf w}\in\mathbb{C}^{N} $ as
\begin{align}
{\bf w}={\bf s}_1+{\bf s}_2
\end{align}
where $ {\bf s}_1\in\mathcal{R}([{\bf a}(\theta_0),{\bf a}(\theta_k)]) $,
$ {\bf s}_2\in\mathcal{R}^{\bot}([{\bf a}(\theta_0),{\bf a}(\theta_k)]) $ and
$ {\bf s}^{\mH}_1{\bf s}_2=0 $.
Notice that 
\begin{align}
{\bf w}^{\mH}{\bf a}(\theta_k)={\bf s}^{\mH}_1{\bf a}(\theta_k),~~
\|{\bf w}\|^2_2=
\|{\bf s}_{1}\|^2_2+\|{\bf s}_{2}\|^2_2.
\end{align}
Clearly, to maximize the utility function in \eqref{eqn0102},
one shall set $ {\bf s}_2 $ as zero.
This completes the proof of \eqref{key005}.

{\it Step 2:} In the second step, we will show that
\begin{align}\label{key010}
\mathcal{R}([{\bf a}(\theta_0),{\bf a}(\theta_k)])=
\mathcal{R}([{\bf{w}}_{\bot},{\bf{w}}_{\Arrowvert}]).
\end{align}
To see this, we recall the expressions of $ {\bf{w}}_{\bot} $
and $ {\bf{w}}_{\Arrowvert} $ in \eqref{word01} and obtain that
\begin{align}\label{eqn55}
[{\bf{w}}_{\bot},{\bf{w}}_{\Arrowvert}]=
[{\bf a}(\theta_0),{\bf a}(\theta_k)]
{\bf T}
\end{align}
where $ {\bf T} $ is given by
\begin{align}
{\bf T}=\begin{bmatrix}
1&0\\
-\dfrac{{\bf a}^{\mH}(\theta_k){\bf a}(\theta_0)}{\|{\bf a}(\theta_k)\|^2_2}&
\dfrac{{\bf a}^{\mH}(\theta_k){\bf a}(\theta_0)}{\|{\bf a}(\theta_k)\|^2_2}
\end{bmatrix}.
\end{align}
If $ {{\bf a}^{\mH}(\theta_k){\bf a}(\theta_0)}\neq 0 $, one can
see that $ {\bf T} $ is
invertible. Thus, Eqn. \eqref{eqn55}
indicates that $ [{\bf{w}}_{\bot},{\bf{w}}_{\Arrowvert}] $ spans the same
column space as that of $ [{\bf a}(\theta_0),{\bf a}(\theta_k)] $. This
completes the proof of \eqref{key010}.

{\it Step 3:} In the third step, we will complete the proof of Proposition 3 by showing that
\begin{align}\label{e065}
{\bf w}_{\triangleright}=c\begin{bmatrix}
{\bf{w}}_{\bot}&{\bf{w}}_{\Arrowvert}\end{bmatrix}
\begin{bmatrix}1& {\beta}_{k,\star}\end{bmatrix}^{\mT},~c\neq 0.
\end{align}
Toward this end, we recall \eqref{key005} and reformulate
problem \eqref{eqn011} as
\begin{subequations}\label{eq2n011}
	\begin{align}
	\label{eqn01202}
	\max_{c_1,c_2\in\mathbb{C}}&~~~\dfrac{|{\bf w}^{\mH}{\bf a}(\theta_0)|^2}
	{\|{\bf w}\|^2_2}\\
	\label{eq2n012}
	{\rm subject~to}&~~~{|{\bf w}^{\mH}{\bf a}(\theta_k)|^2}/
	{|{\bf w}^{\mH}{\bf a}(\theta_0)|^2}={\rho_k}\\
	&~~~{\bf w}=c_1{\bf a}(\theta_0)+c_2{\bf a}(\theta_k).
	\end{align}
\end{subequations}
It should be pointed out that $ c_1\neq 0 $, otherwise, we can obtain
$ {\rho}_k={|{\bf w}^{\mH}_{\triangleright}{\bf a}(\theta_k)|^2}/
{|{\bf w}^{\mH}_{\triangleright}{\bf a}(\theta_0)|^2}={|{\bf a}^{\mH}(\theta_k){\bf a}(\theta_k)|^2}/
{|{\bf a}^{\mH}(\theta_k){\bf a}(\theta_0)|^2}={\widetilde{\rho}_k}>1 $, which is
in contraction with the setting of $ \rho_k\leq 1 $, see
{\it Remark 1}.
According to \eqref{eqn55}, 
$ {\bf w}=c_1{\bf a}(\theta_0)+c_2{\bf a}(\theta_k) $ can be equivalently expressed as
\begin{align}
{\bf w}=[{\bf{w}}_{\bot},{\bf{w}}_{\Arrowvert}]{\bf T}^{-1}
[c_1,~c_2]^{\mT}=c_1[{\bf{w}}_{\bot},{\bf{w}}_{\Arrowvert}][1,~\gamma]^{\mT}
\end{align}
where $ \gamma $ is a specific parameter determined by $ c_1 $, $ c_2 $,
$ {\bf a}(\theta_0) $ and $ {\bf a}(\theta_k) $.
Then, instead of solving problem \eqref{eq2n011} or the original problem 
\eqref{eqn011}, we can formulate the following 
problem:
\begin{subequations}\label{eq23011}
	\begin{align}
	\label{eqn03202}
	\max_{\gamma\in\mathbb{C}}&~~~\dfrac{|{\bf w}^{\mH}{\bf a}(\theta_0)|^2}
	{\|{\bf w}\|^2_2}\\
	\label{eq23012}
	{\rm subject~to}&~~~{|{\bf w}^{\mH}{\bf a}(\theta_k)|^2}/
	{|{\bf w}^{\mH}{\bf a}(\theta_0)|^2}={\rho_k}\\
	&~~~{\bf w}=[{\bf{w}}_{\bot},{\bf{w}}_{\Arrowvert}][1,~\gamma]^{\mT}
	\end{align}
\end{subequations}
which is equivalent to problem \eqref{criterion}.
Combining the result of Proposition 2, we know that
the resulting $ {\bf{w}}_{k} $ in \eqref{e06} is the optimal solution of problem \eqref{eqn011}
under specific prerequisites.
This completes the proof of Proposition 3.

\section{Proof of Proposition 4}
To begin with, we first proof:
\begin{align}
T({\beta}_{k,\star})={\mu}_{k,s}\Leftrightarrow
{\bf c}_{\beta_k}(2)=0 ~{\rm and}~{\bf c}_{\beta_k}(1)\in[0,1].
\end{align}
Toward this end, 
we define $ \widehat{\beta}_{k}\triangleq T^{-1}({\mu}_{k,s}) $.
Then, it's not hard to obtain that
\begin{subequations}\label{e2138}
\begin{align}
{\mu_{k,s}}&={\rm arg}~\min_{{\mu_k}\in{\mathbb{C}_{\mu_k}}}~|\mu_k|\Leftrightarrow\\
T(\widehat{\beta}_{k})&={\rm arg}~\min_{T({\beta_k})\in{\mathbb{C}_{\mu_k}}}~|T({\beta_k})|\Leftrightarrow\\
\widehat{\beta}_{k}&={\rm arg}~\min_{{\beta_k}\in{{\mathbb{C}_{{\beta}_k}}}}~|T({\beta_k})|\Leftrightarrow\\
\label{e2140}\widehat{\beta}_{k}&={\rm arg}~\min_{{\beta_k}\in{{\mathbb{C}_{{\beta}_k}}}}~|\beta_k-1|.
\end{align}	
\end{subequations}
Recalling Proposition 2, $ \beta_{k,\star} $ satisfies
\begin{align}\label{e2139}
{\beta_{k,\star}}={\rm arg}~\max_{{\beta_k}\in{{\mathbb{C}_{{\beta}_k}}}}~|\beta_k|.
\end{align}	
Combining \eqref{e2138} and \eqref{e2139}, one learns that 
$ {\mu}_{k,s}=T({\beta}_{k,\star}) $ or equivalently $ \widehat{\beta}_{k}={\beta}_{k,\star} $
if and only if
\begin{align}\label{e2041}
{\rm arg}~\min_{{\beta_k}\in{{\mathbb{C}_{{\beta}_k}}}}~|\beta_k-1|
={\rm arg}~\max_{{\beta_k}\in{{\mathbb{C}_{{\beta}_k}}}}~|\beta_k|.
\end{align}
Recalling Proposition 1 and Fig. \ref{tspcirclebetabasic}, 
we know that
\eqref{e2041} is true if and only the following two conditions are satisfied.

The first condition is that
\begin{align}\label{e2042}
{\bf c}_{\beta_k}(2)=0.
\end{align}
This is because that the optimal value of the right side of \eqref{e2041} locates
in the intersection of the circle $ \mathbb{C}_{\beta_k} $ and the line that 
passes $ {\bf c}_{\beta_k} $ and $ \bf O $.
And the optimal solution of the left side of \eqref{e2041} locates
in the intersection of $ \mathbb{C}_{\beta_k} $ and the line that 
passes $ {\bf c}_{\beta_k} $ and the point $ [1,0]^{\mT} $.
To obtain an unique $ \beta_k $ that simultaneously optimizes the both sides
of \eqref{e2041}, the three points, i.e., $ \bf O $, $ {\bf c}_{\beta_k} $ and 
$ [1,0]^{\mT} $ must be colinear, i.e.,
$ {\bf c}_{\beta_k}(2)=0 $, otherwise, \eqref{e2041} cannot be true.

On the basis of the above condition (i.e., \eqref{e2042}), the second condition that
makes \eqref{e2041} satisfied is
\begin{align}\label{e2043}
{\bf c}_{\beta_k}(1)\in[0,1].
\end{align}
One can readily verify that \eqref{e2041} holds true
only if both \eqref{e2042} and \eqref{e2043}
are true.

On the contrary, it can be similarly validated that \eqref{e2041} is satisfied,
provided that
\eqref{e2042} and \eqref{e2043} are true.
To summarize, we learn that 
$ T({\beta}_{k,\star})={\mu}_{k,s}={\mu}_{k,\star} $ if and only if
\eqref{e2042} and \eqref{e2043} are satisfied.
The derivation of $ T({\beta}_{k,\star})={\mu}_{k,l} $ is similar, 
we omit it for the sake of space limitation.

In addition, if $ {\bf c}_{\beta_k}(2)=0 $, $ \angle g({\bf c}_{\beta_k}) $ would equal to
$ 0 $ or $ \pi $. Combining the expression of $ \beta_{k,\star} $ in \eqref{e2126},
one can readily obtain that $ {\beta_{k,\star}}\in\mathbb{R} $, provided that
$ T({\beta}_{k,\star})\in\{{\mu}_{k,s},{\mu}_{k,l}\} $.
This completes the proof of Proposition 4.
\section{Proof of Corollary 1}
We divide the proof derivation into three steps.

{\it Step 1}: In the first step, we will proof that 
\begin{align}\label{eqn115}
{\bf B}_k(1,2)\in\mathbb{R}~~{\rm and}~~{\bf B}_k(1,2)\leq 0
\end{align}
provided that $ {\bf w}_{k-1}={\bf a}(\theta_0) $.
To do so, we substitute $ {\bf w}_{k-1} $ into $ {\bf B}_k(1,2) $ and obtain
that
\begin{align}\label{eqn116}
\!\!\!\!\!\dfrac{{\bf B}_k(1,2)}{-\rho_k}&=
{\bf{w}}^{\mH}_{\bot}{\bf{a}}(\theta_0)
{\bf{a}}^{\mH}(\theta_0){\bf{w}}_{\Arrowvert}\nonumber\\
&=({\bf w}_{k-1}-{\bf{w}}_{\Arrowvert})^{\mH}{\bf{a}}(\theta_0)
{\bf{a}}^{\mH}(\theta_0){\bf{w}}_{\Arrowvert}\nonumber\\
&=
\dfrac{\|{\bf a}(\theta_0)\|^2_2\cdot|{\bf{a}}^{\mH}(\theta_0){\bf{a}}(\theta_k)|^2}
{\|{\bf a}(\theta_k)\|^2_2}\!-\!
\dfrac{|{\bf{a}}^{\mH}(\theta_0){\bf{a}}(\theta_k)|^4}
{\|{\bf a}(\theta_k)\|^4_2}
\end{align}
which indicates that $ {\bf B}_k(1,2)\in\mathbb{R} $.
Moreover, from \eqref{eqn116}, one can see that $ {\bf B}_k(1,2)\leq 0 $ is
equivalent to
\begin{align}\label{eqn117}
\dfrac{\|{\bf a}(\theta_0)\|^2_2\cdot|{\bf{a}}^{\mH}(\theta_0){\bf{a}}(\theta_k)|^2}
{\|{\bf a}(\theta_k)\|^2_2}\geq
\dfrac{|{\bf{a}}^{\mH}(\theta_0){\bf{a}}(\theta_k)|^4}
{\|{\bf a}(\theta_k)\|^4_2}
\end{align}
or simply
\begin{align}\label{eqn125}
{\|{\bf a}(\theta_0)\|^2_2\cdot\|{\bf a}(\theta_k)\|^2_2}
\geq
{|{\bf{a}}^{\mH}(\theta_0){\bf{a}}(\theta_k)|^2}.
\end{align}
Since \eqref{eqn125} is a direct result of Cauchy-Schwarz inequality, we
know $ {\bf B}_k(1,2)\leq 0 $ is true.
Thus, the proof of \eqref{eqn115} is completed.

{\it Step 2}: In this step, we will show that if $ \rho_k< {\widetilde{\rho}_k} $, then
\begin{align}\label{pro1}
{\bf B}_k(2,2)> 0.
\end{align}
To do so, we recall the expression of $ {\bf B}_k(2,2) $
and convert \eqref{pro1} as
\begin{align}\label{pro2}
| {\bf{w}}^{\mH}_{\Arrowvert}{\bf{a}}(\theta_k) |^2>
\rho_k| {\bf{w}}^{\mH}_{\Arrowvert}{\bf{a}}(\theta_0) |^2.
\end{align}
Since
\begin{align}\label{pro3}
{\bf{w}}_{\Arrowvert}=\dfrac{{\bf a}(\theta_k){\bf a}^{\mH}(\theta_k){\bf w}_{k-1}}
{{\bf a}^{\mH}(\theta_k){\bf a}(\theta_k)}
\end{align}
one can further rewrite Eqn. \eqref{pro2} as
\begin{align}\label{pro4}
\rho_k< \dfrac{| {\bf{w}}^{\mH}_{\Arrowvert}{\bf{a}}(\theta_k) |^2}
{| {\bf{w}}^{\mH}_{\Arrowvert}{\bf{a}}(\theta_0) |^2}=\dfrac
{\left|
	\frac{{\bf w}^{\mH}_{k-1}{\bf a}(\theta_k){\bf a}^{\mH}(\theta_k)}
	{{\bf a}^{\mH}(\theta_k){\bf a}(\theta_k)}{\bf{a}}(\theta_k)	
	\right|^2}
{\left|
	\frac{{\bf w}^{\mH}_{k-1}{\bf a}(\theta_k){\bf a}^{\mH}(\theta_k)}
	{{\bf a}^{\mH}(\theta_k){\bf a}(\theta_k)}{\bf{a}}(\theta_0)
	\right|^2}={\widetilde{\rho}_k}.
\end{align}
Clearly, 
if $ \rho_k<{\widetilde{\rho}_k} $, it is readily known that \eqref{pro1} holds true.

Now recalling the expression of $ {\bf c}_{\beta_k} $ in \eqref{eqn0030}, 
and combining the conclusions of {\it Step 1}
and {\it Step 2}, one obtains that
\begin{align}\label{e2175}
{\bf c}_{\beta_k}(2)=0~~{\rm and}~~
{\bf c}_{\beta_k}(1)\geq 0
\end{align}
provided that $ {\bf w}_{k-1}={\bf a}(\theta_0) $ and 
$ 0\leq\rho_k<{\widetilde{\rho}_k} $.

{\it Step 3}: In the third step, we will show that
\begin{align}\label{pro5}
{\bf c}_{\beta_k}(1)\leq 1
\end{align}
provided that $ {\bf w}_{k-1}={\bf a}(\theta_0) $, 
$ 0\leq\rho_k\leq\min\left\{{\widetilde{\rho}_k},\overline{{\rho}}_k\right\} $
and $ \rho_k\neq{\widetilde{\rho}_k} $.

To do so, we first note that 
$ \rho_k<{\widetilde{\rho}_k} $ indicates $ {\bf B}_k(2,2)>0 $.
Then,
$ {\bf c}_{\beta_k}(1)\leq 1 $ can be rewritten as
\begin{align}\label{e01}
-{\Re}({\bf B}_k(1,2))=-{\bf B}_k(1,2)\leq{\bf B}_k(2,2)
\end{align}
where $ {\bf B}_k(1,2)\in\mathbb{R} $ has been utilized.
Recalling \eqref{eqn116} and the expression of $ {\bf B}_k(2,2) $,
one can reshape \eqref{e01} as
\begin{align}\label{e02}
{\rho_k}{\bf w}^{\mH}_{k-1}{\bf{a}}(\theta_0)
{\bf{a}}^{\mH}(\theta_0){\bf{w}}_{\Arrowvert}
\leq|{\bf{w}}^{\mH}_{\Arrowvert}{\bf{a}}(\theta_k)|^2
\end{align}
and further
\begin{align}\label{e03}
\dfrac{{\rho_k}\|{\bf a}(\theta_0)\|^2_2\cdot|{\bf{a}}^{\mH}(\theta_0){\bf{a}}(\theta_k)|^2}
{\|{\bf a}(\theta_k)\|^2_2}\leq|{\bf{a}}^{\mH}(\theta_0){\bf{a}}(\theta_k)|^2.
\end{align}
In fact, \eqref{e03} can be reformulated as
$ {\rho_k}\leq\overline{{\rho}}_k $. 
As a consequence, we learn that
$ {\bf c}_{\beta_k}(1)\leq 1 $,
provided that $ {\bf w}_{k-1}={\bf a}(\theta_0) $, 
$ 0\leq\rho_k\leq\min\left\{{\widetilde{\rho}_k},\overline{{\rho}}_k\right\} $
and $ \rho_k\neq{\widetilde{\rho}_k} $.

Recalling \eqref{e2175}, it can be summarized that 
$ {\bf c}_{\beta_k}(2)=0 $ and $ {\bf c}_{\beta_k}(1)\in[0,1] $,
provided that $ {\bf w}_{k-1}={\bf a}(\theta_0) $, 
$ 0\leq\rho_k\leq\min\left\{{\widetilde{\rho}_k},\overline{{\rho}}_k\right\} $
and $ \rho_k\neq{\widetilde{\rho}_k} $.
According to Proposition 4, we have $ T({\beta}_{k,\star})={\mu}_{k,s}={\mu}_{k,\star} $, which implies that $ \textrm{C}^2\textrm{-WORD} $ obtains the same weight
vector $ {\bf w}_{k} $ as ${\textrm A}^2{\textrm{RC}}$. This completes the proof.

\section{Proof of Corollary 2}
Before the proof, we give the following property $ \textbf{P} $ about set $ \mathbb{V} $, which is useful in the
later derivations.
\begin{align}
\boxed{\textbf{P}:
{\rm For}~\forall{\bf v}_1,{\bf v}_2\in\mathbb{V},
\forall c_1,c_2\in\mathbb{R}, {\rm we~ have~} c_1{\bf v}_1+c_2{\bf v}_2\in\mathbb{V}.
}\nonumber
\end{align}
The above property $ \textbf{P} $ can be readily proofed from the definition of set $ \mathbb{V} $, we omit the details
for space limitation.


Then, it requires three steps to complete the proof of Corollary 2.

{\it Step 1}: In the first step, we will proof that
\begin{align}\label{eqn001}
{\Im}({\bf v}^{\mH}_1{\bf v}_2)=0,~\forall {\bf v}_1,{\bf v}_2\in{\mathbb{V}}.
\end{align}
Without loss of generality, we assume that $ N $ is even, the case that $ N $ is odd is similar.
Then for $ \forall{\bf v}\in\mathbb{V} $, we denote an associated vector $ {\bf d} $ by
\begin{equation}\label{ula0004}
{\bf d}\triangleq
\begin{bmatrix}
{\bf v}(1)& {\bf v}(2)& \cdots& {\bf v}(N/2)
\end{bmatrix}^{\mT}.
\end{equation}
Recalling the definition of set $ \mathbb{V} $, we know that
\begin{align}\label{ula0034}
\begin{bmatrix}
{\Re}({\bf v})\\ 
{\Im}({\bf v})
\end{bmatrix}
={\bf G}_1\begin{bmatrix}
{\Re}({\bf d})\\ 
{\Im}({\bf d})
\end{bmatrix},~~
\begin{bmatrix}
{\Im}({\bf v})\\ 
-{\Re}({\bf v})
\end{bmatrix}
={\bf G}_2\begin{bmatrix}
{\Re}({\bf d})\\ 
{\Im}({\bf d})
\end{bmatrix}
\end{align}
where $ {\bf G}_1 $ and $ {\bf G}_2 $ are defined as
\begin{align}\label{G12}
{\bf G}_1\triangleq\begin{bmatrix}
{\bf I}& {\bf 0}\\ 
{\bf J}& {\bf 0} \\ 
{\bf 0}& {\bf I} \\ 
{\bf 0}& -{\bf J} 
\end{bmatrix},~~
{\bf G}_2\triangleq\begin{bmatrix}
{\bf 0}& {\bf I}\\ 
{\bf 0}& -{\bf J} \\ 
-{\bf I}& {\bf 0} \\ 
-{\bf J}& {\bf 0}
\end{bmatrix} 
\end{align}
with $ {\bf I} $ standing for the $ N/2 $-dimensional identity matrix,
$ {\bf J} $ denoting the $ N/2 $-dimensional exchange matrix with ones on its anti-diagonal and
zeros elsewhere.
It can be validated that $ {\bf G}^{\mT}_1{\bf G}_2={\bf 0} $.
Then for $ \forall {\bf v}_1,{\bf v}_2\in\mathbb{V} $, we have
\begin{align}\label{ula0021}
{\Im}({\bf v}^{\mH}_1{\bf v}_2)=
\begin{bmatrix}
{\Re}({\bf d}^{\mT}_{1})& 
{\Im}({\bf d}^{\mT}_{1})
\end{bmatrix}{\bf G}^{\mT}_1
{\bf G}_2	
\begin{bmatrix}
{\Re}({\bf d}_{2})\\ 
{\Im}({\bf d}_{2})
\end{bmatrix}=0
\end{align}	
where $ {\bf d}_1 $ and $ {\bf d}_2 $ stand for the associated vectors of $ {\bf v}_1 $ and
$ {\bf v}_2 $, respectively.
This completes the proof of \eqref{eqn001}.

{\it Step 2}: In this step, it is necessary to show that if $ 0\leq\rho_k<{\widetilde{\rho}_k} $, then
$ {\bf B}_k(2,2)> 0 $. See {\it Step 2} in the derivation of Corollary 1
for details.

{\it Step 3}: In this step, we first assume that $ {\bf w}_{k-1}\in\mathbb{V} $.
Then considering the centro-symmetric arrays
(both $ {\bf a}(\theta_0) $ and $ {\bf a}(\theta_k) 
$ are in set $ \mathbb{V} $), we have
$ {\bf{a}}^{\mH}(\theta_k){\bf{w}}_{k-1}\in\mathbb{R} $.
Recalling the property $ \textbf{P} $,
one obtains
$ {\bf{w}}_{\Arrowvert}=
{\bf{a}}(\theta_k){\bf{a}}^{\mH}(\theta_k){\bf{w}}_{k-1}/\|{\bf a}(\theta_k)\|^2_2\in\mathbb{V} $ 
and $ {\bf{w}}_{\bot}={\bf{w}}_{k-1}-{\bf{w}}_{\Arrowvert}\in\mathbb{V} $.
Therefore, we have $ {\Im}({\bf{w}}^{\mH}_{\bot}{\bf{a}}(\theta_0)
{\bf{a}}^{\mH}(\theta_0){\bf{w}}_{\Arrowvert})=0 $ and
\begin{align}\label{pro7}
{\bf c}_{\beta_k}(2)&={\Im}({\bf B}_k(1,2))/{\bf B}_{k}(2,2)\nonumber\\
&=-\rho_k{\Im}({\bf{w}}^{\mH}_{\bot}{\bf{a}}(\theta_0)
{\bf{a}}^{\mH}(\theta_0){\bf{w}}_{\Arrowvert})/{\bf B}_{k}(2,2)=0.
\end{align}
Combining the result of {\it Step 2} (more exactly, inequity \eqref{pro1}),
we have
\begin{align}\label{pro8}
{\bf c}_{\beta_k}(1)&=-{\Re}({\bf B}_k(1,2))/{\bf B}_{k}(2,2)\nonumber\\
&=\rho_k{\bf{w}}^{\mH}_{\bot}{\bf{a}}(\theta_0)
{\bf{a}}^{\mH}(\theta_0){\bf{w}}_{\Arrowvert}/{\bf B}_{k}(2,2)\geq 0
\end{align}
provided that $ {\bf{w}}^{\mH}_{\bot}{\bf{a}}(\theta_0)
{\bf{a}}^{\mH}(\theta_0){\bf{w}}_{\Arrowvert}\geq 0 $ and
$ 0\leq\rho_k\leq {\widetilde{\rho}_k} $.

Also, 
it can be observed from
$ {\bf{w}}^{\mH}_{\bot}{\bf{a}}(\theta_0)
{\bf{a}}^{\mH}(\theta_0){\bf{w}}_{\Arrowvert}\geq 0 $ that
$ {\bf{w}}^{\mH}_{k-1}{\bf{a}}(\theta_0)
{\bf{a}}^{\mH}(\theta_0){\bf{w}}_{\Arrowvert}\geq
{\bf{w}}^{\mH}_{\Arrowvert}{\bf{a}}(\theta_0)
{\bf{a}}^{\mH}(\theta_0){\bf{w}}_{\Arrowvert} $, and further
\begin{align}\label{pro9}
&~~~\dfrac{{\bf{w}}^{\mH}_{k-1}{\bf a}(\theta_0){\bf a}^{\mH}(\theta_0){\bf a}(\theta_k){\bf a}^{\mH}(\theta_k)
	{\bf{w}}_{k-1}}
{{\bf a}^{\mH}(\theta_k){\bf a}(\theta_k)}\nonumber\\
&={\bf{w}}^{\mH}_{k-1}{\bf a}(\theta_0){\bf a}^{\mH}(\theta_0){\bf{w}}_{\Arrowvert}
\geq{\bf{w}}^{\mH}_{\Arrowvert}{\bf a}(\theta_0){\bf a}^{\mH}(\theta_0){\bf{w}}_{\Arrowvert}
\geq 0.
\end{align}
On this basis, if $ {\bf{w}}^{\mH}_{\bot}{\bf{a}}(\theta_0)
{\bf{a}}^{\mH}(\theta_0){\bf{w}}_{\Arrowvert}\geq 0 $, we have
\begin{align}\label{pr10}
\dfrac{{\bf{w}}^{\mH}_{k-1}{\bf a}(\theta_k){\bf a}^{\mH}(\theta_k){\bf a}(\theta_k){\bf a}^{\mH}(\theta_k)
{\bf{w}}_{k-1}}
{{\bf{w}}^{\mH}_{k-1}{\bf a}(\theta_0){\bf a}^{\mH}(\theta_0){\bf a}(\theta_k){\bf a}^{\mH}(\theta_k)
	{\bf{w}}_{k-1}}={\breve{\rho}_k}
\end{align}
where $ {\breve{\rho}_k} $ has been defined in \eqref{rho2}. If $ {\rho_k}\leq{\breve{\rho}_k} $ is satisfied, then from \eqref{pro9}, \eqref{pr10} and the expression of $ {\bf{w}}_{\Arrowvert} $ in \eqref{pro3}, 
we obtain
\begin{align}
{\rho_k}{\bf{w}}^{\mH}{\bf a}(\theta_0){\bf a}^{\mH}(\theta_0){\bf{w}}_{\Arrowvert}\leq
{\bf{w}}^{\mH}_{\Arrowvert}{\bf a}(\theta_k){\bf a}^{\mH}(\theta_k){\bf{w}}_{\Arrowvert}
\end{align}
which can be reformulated as
$0\leq
{\rho_k}{\bf{w}}^{\mH}_{\bot}{\bf{a}}(\theta_0)
{\bf{a}}^{\mH}(\theta_0){\bf{w}}_{\Arrowvert}\leq| {\bf{w}}^{\mH}_{\Arrowvert}{\bf{a}}(\theta_k) |^2-
\rho_k| {\bf{w}}^{\mH}_{\Arrowvert}{\bf{a}}(\theta_0) |^2
$, or simply $ 0\leq-{\Re}({\bf B}_k(1,2))\leq{\bf B}_{k}(2,2) $.
As a result, one gets
\begin{align}\label{pr12}
{\bf c}_{\beta_k}(1)=-{\Re}({\bf B}_k(1,2))/{\bf B}_{k}(2,2)\leq 1.
\end{align}

Then we can summarize that for a centro-symmetric array, if $ {\bf w}_{k-1}\in\mathbb{V} $,
$ 0\leq\rho_k\leq\min\left\{{\widetilde{\rho}_k},{\breve{\rho}_k}\right\} $,
$ \rho_k\neq{\widetilde{\rho}_k} $ and
$ {\bf{w}}^{\mH}_{\bot}{\bf{a}}(\theta_0)
{\bf{a}}^{\mH}(\theta_0){\bf{w}}_{\Arrowvert}\geq 0 $,
then $ {\bf c}_{\beta_k}(2)=0 $ and $ {\bf c}_{\beta_k}(1)\in[0,1] $.
Recalling Proposition 4, we have $ T({\beta}_{k,\star})={\mu_{k,\star}} $
and $ {\beta}_{k,\star}\in\mathbb{R} $.
Thus, an identical weight vector $ {\bf w}_{k} $ will be resulted by $ \textrm{C}^2\textrm{-WORD} $ and ${\textrm A}^2{\textrm{RC}}$ in the $ k $th response control step.
More importantly, we have $ {\bf w}_{k}={\bf{w}}_{\bot}+
{\beta}_{k,\star}{\bf{w}}_{\Arrowvert}\in\mathbb{V} $.
Consequently, it can be recursive out $ {\bf w}_{k}\in\mathbb{V} $
and $ T({\beta}_{k+1,\star})={\mu_{k+1,\star}} $, provided that
$ {\bf w}_{k-1}\in\mathbb{V} $, and both \eqref{rho1}
and \eqref{rho2} are satisfied. Then obviously, if $ {\bf w}_{0}\in\mathbb{V} $, we can infer that $ T({\beta}_{k,\star})={\mu_{k,\star}} $
(i.e., $ \textrm{C}^2\textrm{-WORD} $ becomes equivalent to ${\textrm A}^2{\textrm{RC}}$) for $ k=1,2\cdots $, 
as long as \eqref{rho1} and \eqref{rho2} hold true for every subscript $ k $.
This completes the proof.

\bibliography{C2WORDPART10718}
\end{document}